\documentclass[aps,superscriptaddress,twocolumn,nofootinbib]{revtex4}
\input epsf

\usepackage{natbib}
\usepackage{graphicx}
\usepackage{dcolumn}
\begin{document}
\title{The long memory of the efficient market}
\author{Fabrizio Lillo}
\affiliation{Santa Fe Institute, 1399 Hyde Park Road, Santa Fe, NM 87501}
\affiliation{INFM and Dipartimento di Fisica e Tecnologie Relative, 
Universit\`a di Palermo, Viale delle Scienze, I-90128 Palermo, Italy.}
\author{J. Doyne Farmer}
\affiliation{Santa Fe Institute, 1399 Hyde Park Road, Santa Fe, NM 87501}
\date{\today}
\begin{abstract}
For the London Stock Exchange we demonstrate that the signs of orders
obey a long-memory process.  The autocorrelation function decays
roughly as $\tau^{-\alpha}$ with $\alpha \approx 0.6$, corresponding
to a Hurst exponent $H \approx 0.7$. This implies that the signs of
future orders are quite predictable from the signs of past orders; all
else being equal, this would suggest a very strong market
inefficiency.  We demonstrate, however, that fluctuations in order
signs are compensated for by anti-correlated fluctuations in
transaction size and liquidity, which are also long-memory processes.
This tends to make the returns whiter. We show that
some institutions display long-range memory and others don't.

\bigskip

\bigskip

{\it Key words:} ~~Long-memory processes, order flow, limit order book, market efficiency.

\medskip 

{\it JEL Classification:}~~C00;C1;G00;G1.
\end{abstract}
\maketitle
\section{Introduction}

Roughly speaking, a random process is said to have long-memory if it
has an autocorrelation function that is not integrable.  This happens,
for example, when the autocorrelation function decays asymptotically
as a power law of the form $\tau^{-\alpha}$ with $\alpha < 1$. This is
important because it implies that values from the distant past can
have a significant effect on the present, and implies anomalous
diffusion in a stochastic process whose increments have long-memory.
We present a more technical discussion of long-memory processes in
Section \ref{longmem}.

In this paper we make use of a data set from the London Stock Exchange
(LSE), which contains a full record of individual orders and
cancellations, to show that stocks in the LSE display a remarkable
long-memory property\footnote{Bouchaud et al. (2004)
  independently discovered the long-memory property of order signs for
  stocks in the Paris exchange.  We thank them for acknowledging the
  oral presentation of our results in May 2003.}.  We label each event
as either a buy or a sell order, and assign it $\pm 1$ accordingly.
The autocorrelation of the resulting time series shows a power law
autocorrelation function, with exponents that are typically about
$0.6$, in the range $0.36 < \alpha < 0.77$.  Positive autocorrelation
coefficients are seen at statistically significant levels over lags of
many thousand events, spanning many days.  Thus the memory of the
market is remarkably long.

This immediately raises a conundrum concerning market efficiency.  All
other things being equal, such strong long-memory behavior would imply
strong predictability of the price return, easily exploitable for
substantial profits.  How can this be compatible with market
efficiency?  We show that this is at least partially solved by other
properties of the market adjusting in order to compensate, making the
market more efficient\footnote{For a discussion of what we mean by
  ``efficient'', see Section~\ref{efficiency}.}.  In particular, the
relative volume of buy and sell market orders, and the relative buy
and sell liquidity, are skewed in opposition to the imbalance in order
signs.  For example, suppose the long-memory of previous order signs
predicts that buy market orders are more likely in the near future.
All things being equal, since buy market orders have a positive price
response, the price should go up.  But the market compensates for
this: When buy orders become more likely, the ratio of buy market
order size to the volume at the best ask tends to be smaller than
normal.  This implies that the probability of a given buy order
penetrating the best price to trigger a positive price change is
smaller than normal.  Similarly, the opposite is true when sell orders
are more likely.  This is at least one of the effects that contributes
to keeping the price returns roughly white.  We demonstrate that
market order volume and liquidity are also long-memory processes; we
suggest this is because they are compensating for changes in the
predictability of order signs in order to keep the market more
efficient.

This brings up the interesting question of what actually causes these
long-memory properties of markets.  While not answering this
question, we provide a clue about the answer by making use of the
institutional codes associated with each order.  Some institutions
show long-memory quite clearly, while others do not.

The paper is organized as follows: In Section~\ref{review} we present
a brief summary of previous work on long-memory processes in
economics, and discuss our work in the context of this literature.  In
Section~\ref{data} we give a summary of the data set, and in
Section~\ref{longmem} we provide a more technical discussion of
long-memory processes and the statistical techniques we use in this
paper.  Then in Section~\ref{statresults} we present the evidence that
sequences of order signs are long-memory processes, and in particular
we demonstrate that these series pass stringent tests so that we can
be sure that they are long-memory with a high degree of confidence.
In Section~\ref{efficiency} we show how other processes compensate for
the long-memory of order signs, tending to keep price changes roughly
efficient, and show that both order size and liquidity are also
long-memory processes.  In Section~\ref{investors} we break down the
orders by institution and show that the behavior of some institutions
shows long-memory quite clearly, while others do not show it at all.
In Section~\ref{implications} we discuss the implications of the
long-memory properties of order flow for delayed market impact.  We
conclude in Section~\ref{conclusion}, discussing some of the broader
issues and the remaining questions.

\section{long-memory processes in economics time series\label{review} }

Long-memory processes have been observed in different natural and
human phenomena ranging from the level of rivers to the temperature of
the Earth (Beran, 1994).  A good survey of the econometric approach to
long-memory is given in Baillie (1996). The range of applications
of long-memory processes in economics spans from macroeconomics
to finance. In macroeconomics, for example, Diebold and Rudebusch
(1989) find evidence of long-memory in the quarterly
post World War II US real GNP data. Even if several criticisms have been
raised to this work, subsequent analyses confirm the evidence of
long-memory properties of GNP data.
Baillie, Chung and Tieslau (1995) find that the monthly US
Consumer Price Index (CPI) inflation time series has long-memory
properties. A related study by Hassler and Wolters (1995)
considers long-memory in inflation. Finally, Shea (1991) and
Backus and Zin (1993) find evidence of long-memory in the
context of the term structure of interest rates.

The study of possible long-memory properties of time series in finance
is even more widespread.  There has been a long-standing debate as to
whether or not asset prices have long-memory properties.  Several
authors have claimed that the time series of stock returns for stock
prices or indices display long-memory (Mandelbrot, 1971, Greene and
Fielitz, 1977).  More recently Lo (1991) re-examined these results and
showed that the statistical R/S test used by Mandelbrot and Green and
Fielitz is too weak and unable to distinguish between long and short
memory (see also section~\ref{longmem} and Campbell et al. 1997)). By
introducing a modified R/S test, Lo concluded that daily stock returns
do not display long-memory properties.  This conclusion has been in
turn criticized by Willinger, Taqqu and Teverovsky (1999).  These
authors showed with numerical simulations that the modified R/S test
leads to the rejection of the null hypothesis of short-memory when
applied to synthetic time series with a low degree of
long-memory. Since financial data tipically display low degree of
long-memory, Willinger, Taqqu and Teverovsky (1999) claim that the
result of Lo is not conclusive. (We will not address this question). 

It is more widely accepted (though still not entirely uncontroversial)
that the volatility of prices is a long-memory process. Specifically
Ding, Granger and Engle (1993) and Breidt, Crato and de Lima (1993)
find evidence of long-memory stochastic volatility in stock returns,
and Harvey (1993) finds evidence for this in exchange rates. These
results led to the development of alternate models for volatility,
such as FIGARCH.  In particular, Baillie, Bollerslev and Mikkelsen
(1996) apply the FIGARCH process to exchange rates, and Bollerslev and
Mikkelsen (1996) apply FIEGARCH, a modification of FIGARCH, to stock
prices.  Another market property that seems to have long-memory
properties is stock market trading volume (Lobato and Velasco, 2000).
Models of long-memory processes include fractional Brownian noise
(Mandelbrot and van Ness, 1968) and the ARFIMA process introduced by
Granger and Joyeux (1980) and Hosking (1981).

In this paper we discover and study the long-memory properties of the
signs of orders in a financial market. Despite the ubiquity of
long-memory processes in economic time series described above, our
result is, to our knowledge, the first demonstration of the presence
of long-memory in a ``microscopic" time series, i.e. a time series
which is not the result of the aggregation of many individual events
(such as the price return). Because of the rapid timescale, we have a
large amount of data, and we are able to demonstrate the existence of
long-memory properties with a very high degree of confidence.  The
order flow is a time series describing the action of trading
institutions, i.e. it is an input to the price formation
mechanism.  From this point of view it is very interesting to see how
the market responds to the existence of a highly predictable
long-memory input, to form prices so as to maintain market efficiency.
Because we are able to see the detailed state of the orderbook, we can
study how bids, offers, and order volumes adjust to compensate, so
that the long-memory properties of order signs and volumes are cancelled
in the directional movements of prices.  We investigate the order
flow mainly in event time. This means that one time step is defined by
the placement of an order of a given type (market order, limit order
and cancellation).  As a consequence the long-memory property we
discuss here is not the effect of inhomogeneous trading during the
day, but rather reflects the strategic way in which traders place
their orders.  Finally, we prove with statistical confidence that the
long-memory property of the sign of the order flow can be present also
at the level of individual trading institution.

\section{Data\label{data}}

In order to a have a representative sample of stocks we select $20$
companies continuously traded at the London Stock Exchange (LSE) in
the 4-year period 1999-2002.  The stocks we analyzed are Astrazeneca
(AZN), Bae Systems (BA.), Baa (BAA), BHP Billiton (BLT), Boots Group
(BOOT), British Sky Broadcasting Group (BSY), Diageo (DGE), Gus (GUS),
Hilton Group (HG.), Lloyds Tsb Group (LLOY), Prudential (PRU), Pearson
(PSON), Rio Tinto (RIO), Rentokil Initial (RTO), Reuters Group (RTR),
Sainsbury (SBRY), Shell Transport \& Trading Co. (SHEL), Tesco (TSCO),
Vodafone Group (VOD), and WPP Group (WPP). Table \ref{summary} gives a
summary of the number of different events for the $20$ stocks.

\begin{table}
\caption{Summary statistics of the $20$ stocks we study for the period
  1999-2002.  The columns give the number of events of each type, in
  thousands.  All events are ``effective'' events -- see the
  discussion in the text.}
\begin{tabular}{l|ccc|r}
tick & market orders & limit orders & cancellations& total\\
\tableline
AZN  &  ~652 & 2,067 & 1,454~~~ &  ~~4,173 \\
BA   &  ~381 &  950 &  598~~~ &  ~~1,929  \\
BAA  &  ~226 &  683 &  487~~~ &  ~~1,397  \\
BLT  &  ~297 &  825 &  557 ~~~&  ~~1,679  \\
BOOT &  ~246 &  711 &  501 ~~~&  ~~1,458  \\
BSY  &  ~404 & 1,120 &  726 ~~~& ~~ 2,250 \\
DGE  &  ~527 & 1,329 &  854 ~~~& ~~ 2,709 \\
GUS  &  ~244 &  734 &  518 ~~~&  ~~1,496  \\
HG.  &  ~228 &  676 &  472 ~~~&  ~~1,377  \\
LLOY &  ~723 & 1,664 & 1,020 ~~~&  ~~3,407 \\
PRU  &  ~448 & 1,227 &  821~~~ &  ~~2,496 \\
PSON &  ~373 & 1,063 &  734 ~~~& ~~ 2,170 \\
RIO  &  ~381 & 1,122 &  771 ~~~&  ~~2,274 \\
RTO  &  ~276 &  620 &  389 ~~~&  ~~1,285  \\
RTR  &  ~479 & 1,250 &  820 ~~~&  ~~2,549 \\
SBRY &  ~284 &  805 &  561 ~~~&  ~~1,650  \\
SHEL &  ~717 & 4,137 & 3,511 ~~~&  ~~8,365 \\
TSCO &  ~471 &  949 &  523 ~~~&  ~~1,943  \\
VOD  & 1,278 & 2,358 & 1,180 ~~~&  ~~4,817 \\
WPP  &  399  & 1,151 &  780 ~~~&  ~~2,330 \\
\tableline
total& 9,034 & 25,441 & 17,277~~ &~51,752 \\
\end{tabular}
\label{summary}
\end{table}

The London Stock Exchange consists of two markets, the electronic
(SETS) exchange, and the upstairs market.  We study only the
electronic exchange.  The data set we analyze contains every action by
every institution participating in this exchange.  In 1999 the
electronic exchange contains roughly $57 \%$ percent of the order flow
for a typical stock, and in $2002$ roughly $62 \%$ percent.  It is
thus always a substantial fraction of the total order flow, and is
believed to be the dominant mechanism for price formation. There are
several types of orders allowed by the exchange, with names such as
``fill or kill" and ``execute or eliminate".  To place our analysis in
more useful terms we label events in terms of their net effect on the
limit order book.  We label any component of an order that results in
an immediate transaction an {\it effective market order}, and any
component of an order that leaves a limit order sitting in the book an
{\it effective limit order}.  A single order may result in multiple
effective orders. For example, consider a crossing limit order, i.e. a
limit order whose limit price crosses the opposing best price
quote. The part of the order that results in an immediate transaction
is counted as an effective market order, while the remaining
non-transacted part (if any) is counted as an effective limit order.
We will call anyone who places effective market orders a {\it
  liquidity taker}, and anyone who places effective limit orders a
{\it liquidity provider}.

We will also lump together any event that results in a queued limit
order being removed without a transaction, and refer to such an event
as a {\it cancellation}.  Henceforth dropping the modifier
``effective'', we can then classify events as one of three types:
market order, limit order and cancellation.  For the set of $20$
stocks described above there is a total of roughly $9$ million market
orders, $25$ million limit orders and $17$ million cancellations.
Throughout this paper, unless otherwise specified, we use the number
of effective events as a measure of time, which we call {\it event
  time}.  We typically do this in terms of the number of events of a
given type, e.g. if we are studying market orders we measure event
time in terms of the number of market orders, and if we are studying
limit orders we measure event time in terms of the number of limit
orders.

Trading begins each day with an opening auction.  There is a period
leading up to the opening auction in which orders are placed but no
transactions take place.  The market is then cleared and for the
remainder of the day (except for occasional exceptional periods) there
is a continuous auction.  We remove all data associated with the
opening auction, and analyze only orders placed during the continuous
auction.

An analysis of the limit order placement shows that in our dataset
approximately $35\%$ of the effective limit orders are placed behind
the best price (i.e. inside the book), $33\%$ are placed at the best
price, and $32\%$ are placed inside the spread.  This is roughly true
for all the stocks except for SHEL, for which the percentages are
$71\%$, $18\%$ and $11\%$, respectively. Moreover for all the stocks
the properties of buy and sell limit orders are approximately the
same.

In our dataset cancellation occurs roughly $32\%$ of the time at the
best price and $68\%$ of the time inside the book. This is quite
consistent across stocks and between the cancellation of buy and sell
limit orders. As for the case of the placement of limit orders, the
only significant deviation is SHEL, for which the percentages are
$14\%$ and $86\%$.

In the following {\it price} will indicate the mid price,
i.e. $p(t)=(a(t)+b(t))/2$ where $a(t)$ and $b(t)$
are the best ask and best bid prices at time $t$,
respectively.

\section{Review of methods for understanding long-memory 
processes\label{longmem}}

\subsection{Definitions of long-memory}

There are several way of characterizing long-memory processes.  A
widespread definition is in terms of the autocovariance function
$\gamma(k)$. We define a process as long-memory if in the limit
$k\to \infty$
\begin{equation}
\gamma(k) \sim k^{-\alpha} L(k)
\label{LMdef}
\end{equation} 
where $0<\alpha<1$ and $L(x)$ is a slowly varying
function\footnote{$L(x)$ is a slowly varying function (see Embrechts et al., 1997)
if $\lim_{x \to \infty} L(tx)/L(x) = 1$.  In the definition above, and
for the purposes of this paper, we are considering only positively
correlated long-memory processes. Negatively correlated
long-memory processes also exist, but the long-memory processes we will
consider in the rest of the paper are all positively correlated.} at
infinity.  The degree of long-memory is given by the
exponent $\alpha$; the smaller $\alpha$, the longer the memory.

Long-memory is also discussed in terms of the Hurst exponent $H$,
which is simply related to $\alpha$.  For a long-memory process
$H=1-\alpha/2$ or $\alpha=2-2H$. Short-memory processes have $H =
1/2$, and the autocorrelation function decays faster than $k^{-1}$.  A
positively correlated long-memory process is characterized by a Hurst
exponent in the interval $(0.5,1)$.  The use of the Hurst exponent is
motivated by the relationship to diffusion properties of the
integrated process.  For normal diffusion, where by definition the
increments do not display long-memory, the standard deviation
asymptotically increases as $t^{1/2}$, whereas for diffusion processes
with long-memory increments, the standard deviation asymptotically
increases as $t^H L(t)$, with $1/2 < H < 1$, and $L(t)$ a slow-varying
function.

Yet another equivalent definition of long-memory dependence can be
given in terms of the behavior of the spectral density for low
frequencies. A long-memory process has a spectral density which
diverges for low frequencies as
\begin{equation}
g(f)\simeq f^{1-2H} L(f),
\end{equation}
where $f$ is the frequency, and $L(f)$ is a slowly varying
function in the limit $f \to 0$.  This follows immediately from
the fact that the autocorrelation and the spectral density are Fourier
transforms of each other.

\subsection{Statistical tests for long-memory}

The empirical determination of the long-memory property of a time
series is a difficult problem. The basic reason for this is that the
strong autocorrelation of long-memory processes makes statistical
fluctuations very large.  Thus tests for long-memory tend to require
large quantities of data and can often give inconclusive results.
Furthermore, different methods of statistical analysis often give contradictory
results.  In this section we review two such tests and discuss some of
their properties.  In particular we discuss the classical R/S test,
which is known to be too weak, and Lo's modified R/S test, which is
known to be too strong.

The basic idea behind the classical R/S test (Hurst, 1951, Mandelbrot,
1972 and 1975) is to compare the minimum and maximum values of running
sums of deviations from the sample mean, renormalized by the sample
standard deviation.  For long-memory processes the deviations are
larger than for non-long memory processes.  The classical R/S test has
been proven to be too weak, i.e. it tends to indicate a time series
has long-memory when it does not.  In fact, Lo (1991) showed that even
for a short-memory process, such as a simple AR(1) process, the
classical R/S test does not reject the null hypothesis of
short-memory.  This fact motivated Lo (1991) to introduce a stronger
test based on a modified R/S statistic.

We now describe Lo's modified R/S test.  Consider a sample time series
$X_1$, $X_2$,...,$X_n$ with sample mean $(1/n)\sum_jX_j$ as $\bar
X_n$.  Let $\hat \sigma^2_x$ and $\hat\gamma_x$
be the sample variance and autocovariance.
The modified rescaled range statistic $Q_n(q)$ is defined by
\begin{eqnarray}
Q_n(q)\equiv ~~~~~~~~~~~~~~~~~~~~~~~~~~~~~~~~~~~~~~~~~~~~~~~~~~~~~~~~~~\\
\equiv \frac{1}{\hat\sigma_n(q)}\left[\max_{1\le k\le n}
\sum_{j=1}^k(X_j-\bar X_n)-\min_{1\le k\le n}
\sum_{j=1}^k(X_j-\bar X_n)\right], \nonumber
\end{eqnarray}
where 
\begin{equation}
\hat\sigma^2_n(q)\equiv \hat \sigma^2_x+2\sum_{j=1}^q\omega_j(q)
\hat\gamma_j,~~~~~~~~~\omega_j(q)\equiv1-\frac{j}{q+1},
\end{equation}
and $q<n$.
It is worth noting that $Q_n(q)$ differs from the classical
R/S statistics of Mandelbrot only in the denominator. 
In the classical R/S test $\hat\sigma_n(q)$ is replaced by
the sample standard deviation $\hat \sigma_x$. 

The optimal value of $q$ to be used in Eq. (3) to compute $Q_n$ must
be chosen carefully. Lo suggested the value $q=[k_n]$, where
\begin{equation}
k_n\equiv\left(\frac{3n}{2}\right)^{\frac{1}{3}}\left(
\frac{2\hat\rho}{1-\hat\rho^2}\right)^{\frac{2}{3}},
\end{equation}
$[k_n]$ indicates the greatest integer less than or equal to
$k_n$ and $\hat \rho$ is the sample first-order autocorrelation
coefficient of the data.  Lo was able to prove that if the process has
finite fourth moment and it has a short-memory dependence 
(and satisfies other supplementary conditions) 
$V_n\equiv Q_n/\sqrt{n}$ tends asymptotically to a random
variable distributed according to 
\begin{equation}
F_V(v)=1+2\sum_{k=1}^{\infty}(1-4k^2v^2)e^{-2(kv)^2}. 
\end{equation}
This result makes
it possible to find the boundaries of a given confidence interval
under the null hypothesis that the time series is short-memory.  When
$V_n$ is outside the interval $[0.809,1.862]$, we can reject the null
hypothesis of short range dependence with $95\%$ confidence.
 
Recently Teverovsky, Taqqu and Willinger (1999) have shown that Lo's
rescaled R/S test is too severe. They showed numerically that
even for a synthetic long-memory time series with a moderate value of
the Hurst exponent (like $H=0.6$) the Lo test cannot reject the null
hypothesis of short range dependence.  For our results here we are
lucky that we are able to pass the rescaled R/S test for long-memory,
but the stringency of this test should be borne in mind in evaluating
our results.

\subsection{Methods of measuring the Hurst exponent}

The determination of the Hurst exponent of a long-memory process is
not an easy task, especially when one cannot make any parametric
assumptions about the investigated time series. Several heuristic
methods have been introduced to estimate the Hurst exponent.  Recently
some authors suggested the use of a ``portfolio" of estimators instead
of relying on a single estimator which could be biased by the property
of the time series under investigation (Taqqu, Teverovsky and Willinger, 1995).
 In this paper we
will discuss four widespread Hurst exponent estimators which we
describe below. These methods are the periodogram method, the R/S
method, Detrended Fluctuation Analysis and the fit of the
autocorrelation function.  We find that the first three methods give
reasonable agreement both in real and in surrogate time series. The
fourth method appears to be more noisy and less reliable.

To use the periodogram method, one first calculates the periodogram
$I(f)$, which is an estimate of the spectral density.
\begin{equation}
I(f)=\frac{1}{2\pi n}\big|\sum_{j=1}^nX_j~e^{ijf}\big|^2
\end{equation} 
where, as before, $n$ is the size of the sample $X_j$. Then a
regression of the logarithm of the periodogram against the logarithm
of $f$ for small values of $f$ gives a slope coefficient that
estimates $1-2H$ (see Eq. 2).  We make our regression on the
lowest $10\%$ of the data (Taqqu, Teverovsky and Willinger, 1995).

The second method is the R/S method (Mandelbrot, 1972 and 1975).  A
description of the method, which is strongly based on R/S statistics,
can be found in Beran (1994).  In summary, we divide a time series of
length $n$ in $K$ blocks of size $n/K$ and we chose logarithmically
spaced values of the lag $k=1,2,4,8...$. For a given value of $k$ we
compute the classical R/S statistics (i.e. Eq.(3) with $\hat \sigma_x$
instead of $\hat \sigma_n(q)$ in the denominator) in each block, by
using the first point as the starting point. When $k<n/K$, one
obtains $K$ different values of the R/S statistics. Finally we plot the value 
of the R/S statistics versus $k$ in double logarithmic scale. The parameter $H$
is obtained by fitting a line to this plot.

The third method is the Detrended Fluctuation Analysis introduced in
Peng et al. (1994).  The time series is first integrated.  The
integrated time series is divided into boxes of equal length $m$.  In
each box, a least squares line is fit to the data (representing the
trend in that box).  The $y$ coordinate of the straight line segments
is denoted by $y_m(k)$.  Next, we detrend the integrated time series,
$y(k)$, by subtracting the local trend, $y_m(k)$, in each box.  The
root-mean-square fluctuation of this integrated and detrended time
series is calculated by
\begin{equation}
F(m)=\sqrt{\frac{1}{n}\sum_{k=1}^n[y(k)-y_m(k)]^2}.
\end{equation}
This computation is repeated over all time scales (box sizes) 
to characterize the relationship between $F(m)$ 
and the box size $m$.  
Typically $F(m)$ will increase with box size $m$. 
The Hurst exponent is obtained by fitting $F(m)$ with a relation
$F(m)\propto m^H$. The proposers of this method claim that it
is able to remove local trends due to bias in the enhanced occurrence 
of a class of events (Peng et al., 1994)

A fourth method is to simply compute the autocorrelation function and
measure $\alpha = 2 - 2H$ by regressing the autocorrelation function
with a power law.  This method, however, suffers from the problem that
the sample errors in adjacent autocorrelation coefficients are
strongly correlated, and so this method is less accurate than the
other two methods discussed above.  Thus, we only use this
method as an indication.  Based on tests on real and surrogate data,
we find that the first three methods all give very similar results; we use
either the R/S method or the periodogram method when we want to get accurate
values of the exponent $\alpha$.

\section{Demonstration of long-memory for order signs\label{statresults}}

\subsection{A quick look at the autocorrelation function}

We consider the symbolic time series obtained in event time by
replacing buy orders with $+1$ and sell orders with $-1$, irrespective of
the volume (number of shares) in the order.  This can be done for
market orders, limit orders, or cancellations.  As we will see, all of
these series show very similar behavior.  We reduce these series to
$\pm 1$ rather than analyzing the signed series of order sizes
$\omega_t$ in order to avoid problems created by the large
fluctuations in order size; analysis of the signed series of order
sizes produces results that do not converge very well.

\begin{figure}[ptb]
\begin{center}
\includegraphics[scale=0.3]{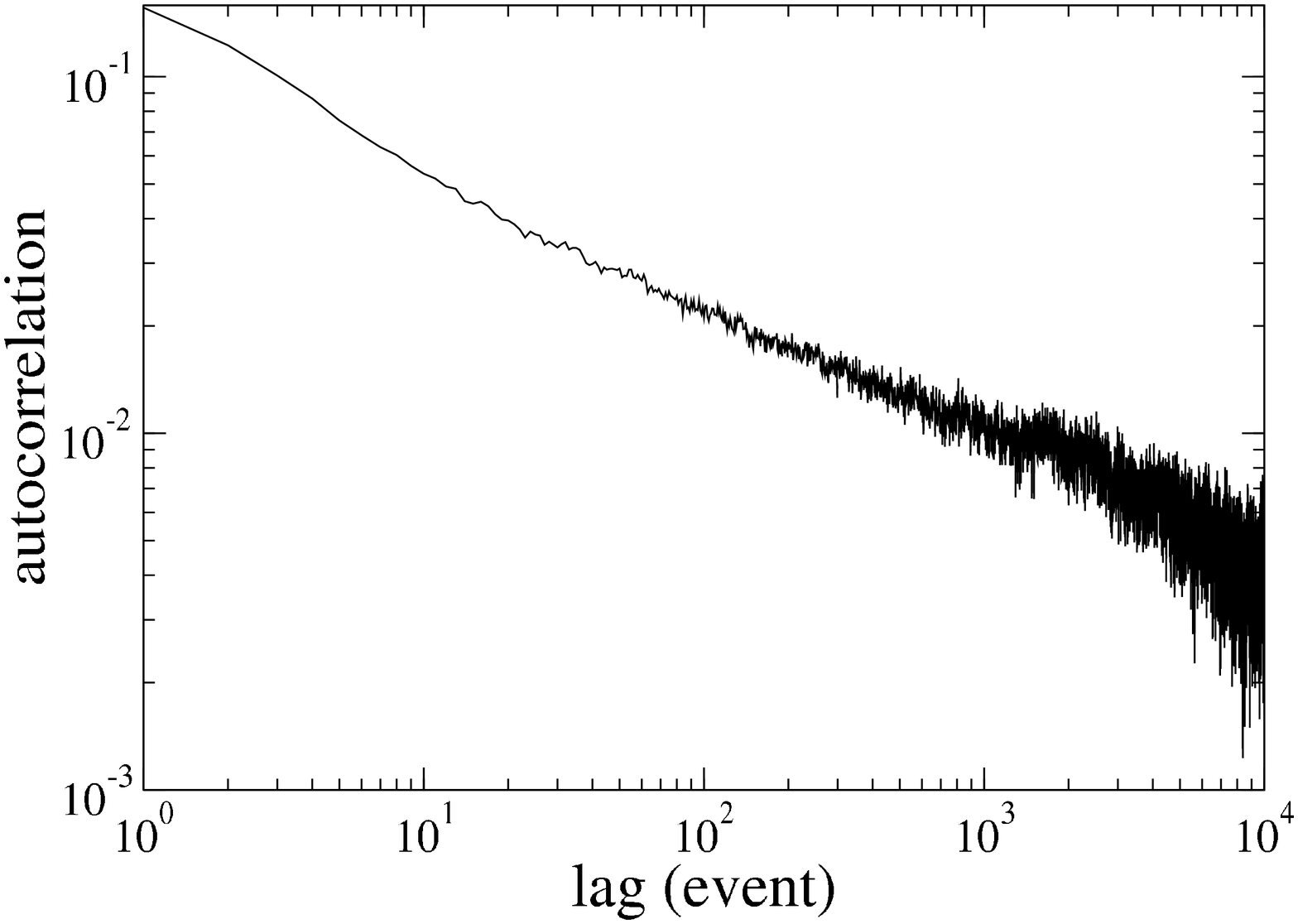}
\includegraphics[scale=0.3]{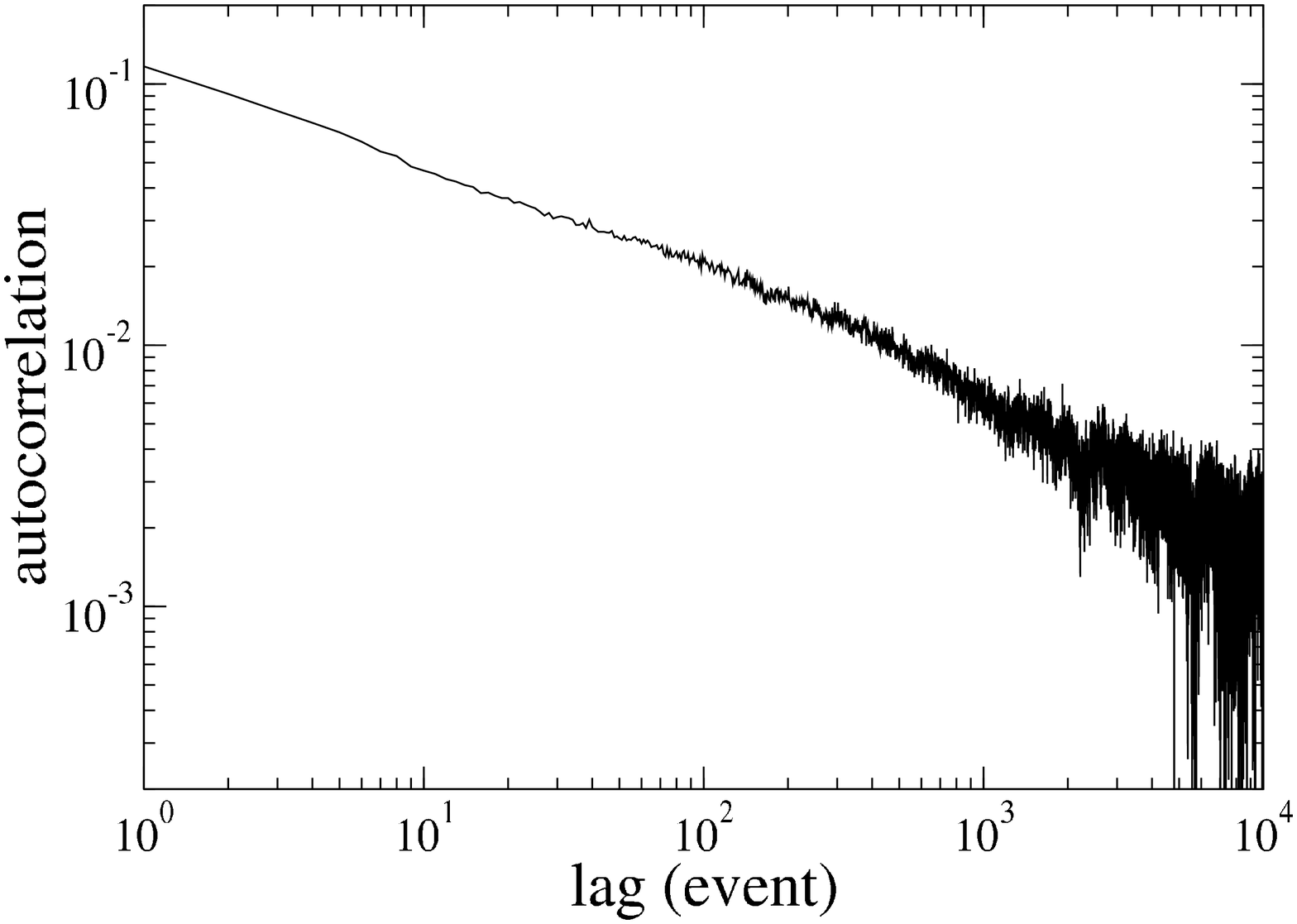}
\includegraphics[scale=0.3]{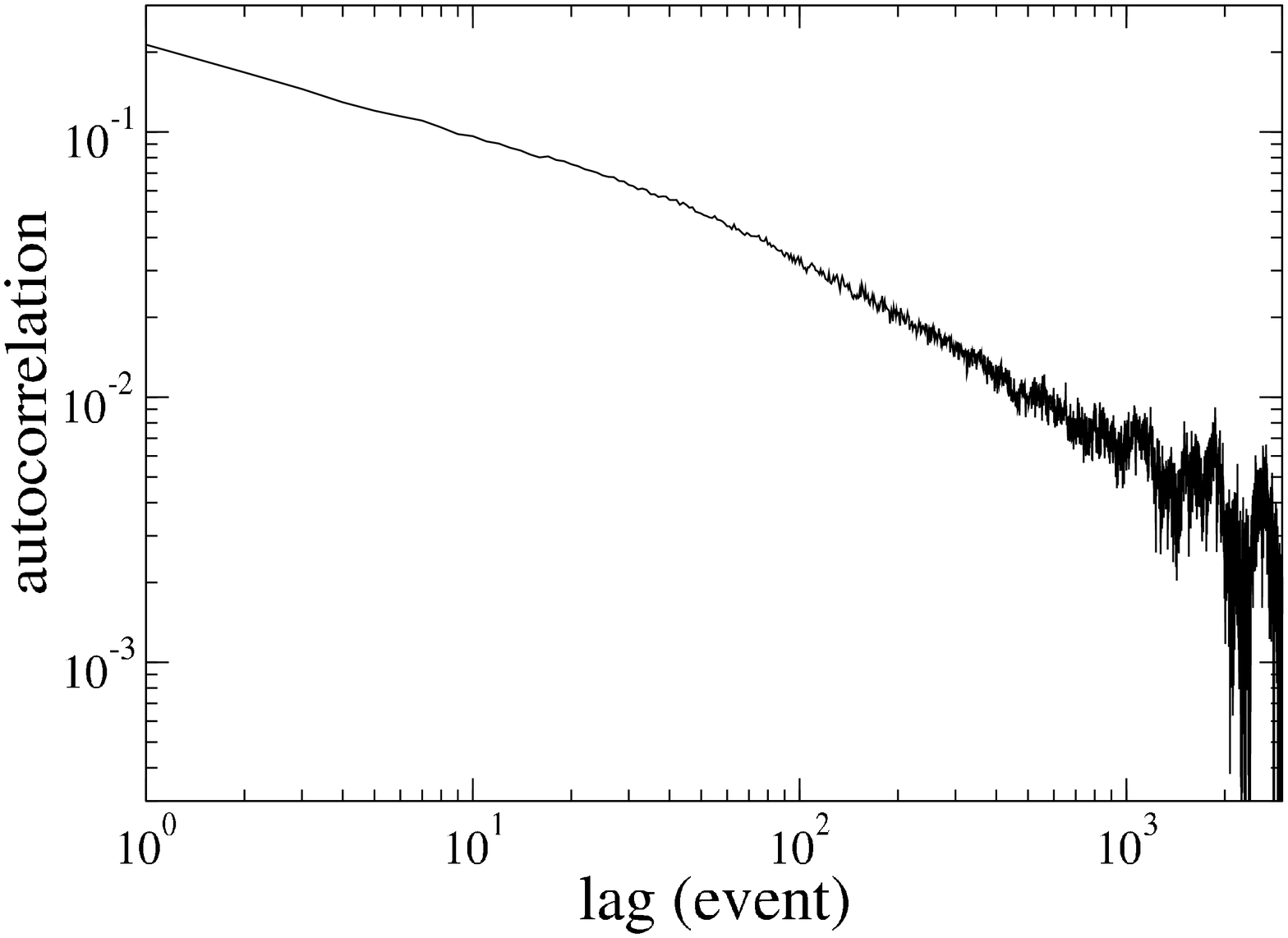}
\caption{Autocorrelation function of sequences of order signs for
Vodafone in the period 1999-2002 in double logarithmic scale for (a)
market orders, (b) limit orders and (c) cancellations.  The lag is
measured in terms of the number of events of each type, e.g., number
of market orders, number of limit orders, etc.  In each case the
autocorrelation function remains positive over periods much longer
than the average number of events in a day.}
\label{VOD-ac-99-02}
\end{center}
\end{figure}

Figure \ref{VOD-ac-99-02}  shows the sample autocorrelation functions of the
order sign time series for Vodafone in the period 1999-2002 in
double logarithmic scale. 
Vodafone was chosen to illustrate the results in this paper because it
is one of the most capitalized and most heavily traded stocks in the
LSE during this period; we see very similar results for all
the other stocks in our dataset. The autocorrelation function for
market orders, limit orders and cancellations decays roughly linearly
over more than $4$ decades, although with some break in the slope for
limit orders and cancellations. This suggests that a power-law
relation $\rho(k) \sim k^{-\alpha}$ is a reasonable asymptotic
approximation for the empirical autocorrelation function. Of course,
for larger lags there are fewer independent intervals, and the
statistical fluctuations are much larger.

Estimating $\alpha$ from the sample autocorrelation using an ordinary
least squares fit gives $\alpha=0.39$ for market orders.  For limit
orders there appears to be a break in the slope, with an exponent
roughly $0.4$ for lags less than roughly $500$ and $0.6$ for larger
lags.  There is a similar break in the slope for cancellations, with a
slope roughly $0.4$ for less than $50$ lags and $0.7$ for larger
lags. As already mentioned, the sample autocorrelation is a poor method
for estimating $\alpha$, and should only be considered an indication;
later on we will use more reliable estimators. But the fact that
$\alpha$ is much smaller than 1 in every case suggests that these
might be long-memory processes.  The memory is quite
persistent, as is evident from the fact that the sample
autocorrelations remain positive over a very long span of time.  The
average daily number of market orders for Vodafone in the investigated
period is approximately $1,300$, whereas the slow decay of the
autocorrelation function in Fig. \ref{VOD-ac-99-02} is seen for lags
as large as $10,000$. This indicates that the long-memory property of
the market order placement is not just an intra-day phenomenon, but
rather spans multiple days, persisting on a timescale of more than a
week.  Similar statements are true for limit orders and cancellations.
See also Section~\ref{realTime}, where we analyze this phenomenon in
real time rather than event time.

\subsection{Statistical evidence for long-memory}

\begin{figure}[ptb]
\begin{center}
\includegraphics[scale=0.3]{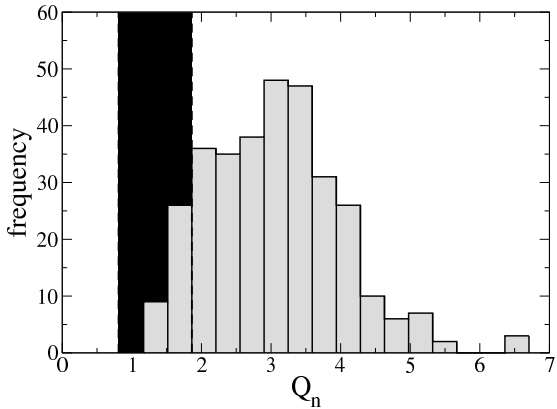}
\caption{Histogram of the statistics of the modified R/S statistics
$Q_n$ for subsets of the market order sign time series. The original
set of $20$ stocks traded in the LSE in the period 1999-2002 is divided
into 324 disjoint subsets as described in the text.  The black region
is the $95\%$ confidence interval region of the null hypothesis of
short-memory.  For $289$ (89\%) of them we can reject the null
hypothesis of long-memory with at least $95\%$ confidence.  Similar
analyses for limit orders and cancellations give even stronger results.}
\label{r_over_s_lo_hist-mo}
\end{center}
\end{figure}

In order to test the presence of long-memory properties in the time
series of market order signs both longitudinally (i.e. analyzing a
stock for different time periods) and cross-sectionally
(i.e. analyzing different stocks) we proceed as follows: We consider
the set of $20$ highly capitalized stocks described in Section 2 for
the 4 year period 1999-2002. Since the number of orders is different
for different stocks in different calendar years, we divide the data
for each year and for each stock into subsets in such a way that each
set contains roughly a fixed number of orders\footnote{The number of
  market orders ranges from $26,438$ for WPP in 1999 to $415,392$ for
  VOD in 2002, the number of limit orders ranges from $51,798$ in 1999
  for BLT to $2,552,410$ for SHEL in 2002, and the number of
  cancellations ranges from $29,395$ for BLT in 1999 to $2,259,526$
  for SHEL in 2002.  We thus divided the market orders into $324$
  subsets ranging in size from $25,000$ to $49,999$; we divided limit
  orders into $468$ subsets ranging in size from $50,000$ to $99,999$,
  and we divided cancellations into $558$ subsets in size ranging from
  $29,000$ to $57,999$.}. To each set we apply the Lo test based on
modified R/S statistics, obtaining a value for the statistics
$Q_n$. Since our time series consists of $+1$ and $-1$ we do not have
problems with the existence of moments. Figure
\ref{r_over_s_lo_hist-mo} shows the histogram of the $324$ values of
$Q_n$ for the subsets of market orders.
For $289$ ($89.2\%$) subsets we can reject the null hypothesis of
short-memory processes with $95\%$ confidence.  Repeating this test
for limit orders and cancellations gives even
stronger results: For limit orders, based on $468$ subsets the
short-memory hypothesis is rejected at the $95\%$ level in $97\%$ of
the cases, and for cancellations using $558$ subsets it is rejected
in $96\%$ of the cases.  We can therefore conclude that these order
sign time series are almost certainly long-memory processes. This result
is even stronger when one considers the severity of this test, as
pointed out in Teverovsky, Taqqu and Willinger (1999).

We have performed a similar analysis for the NYSE, using the Lee and
Ready algorithm to sign the trades (Lee and Ready, 1991).  Despite
some technical problems associated with classifying the trades, it is
quite clear that this is also a long-memory process\footnote{The Lee
  and Ready algorithm is not completely reliable in classifying
  trades.  Fifteen percent of the trades remain unclassified.
  By random substitution of the sign of the unclassified
  trades, it is clear that for NYSE stocks the market
  order sign is a long-memory process with exponents similar to those
  observed in the LSE.}.

\subsection{Estimating the Hurst exponents}

Now that we have established that these are long-memory processes we
determine the Hurst exponent $H$ to see if there is consistency in the
exponent in different years and for different stocks.  Recall that for
a long-memory process the Hurst exponent is related to the exponent
$\alpha$ of the autocorrelation function through $\alpha=2-2H$.
 
\begin{figure}[ptb]
\begin{center}
\includegraphics[scale=0.3]{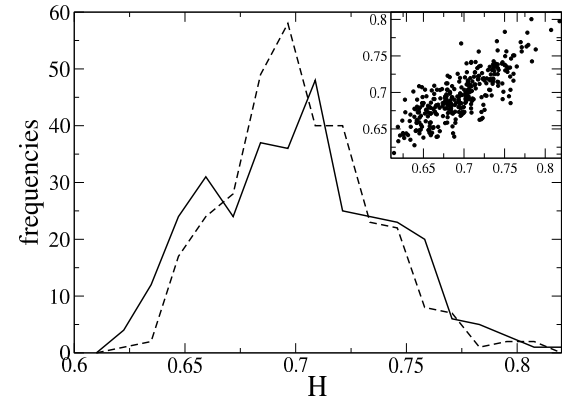}
\includegraphics[scale=0.3]{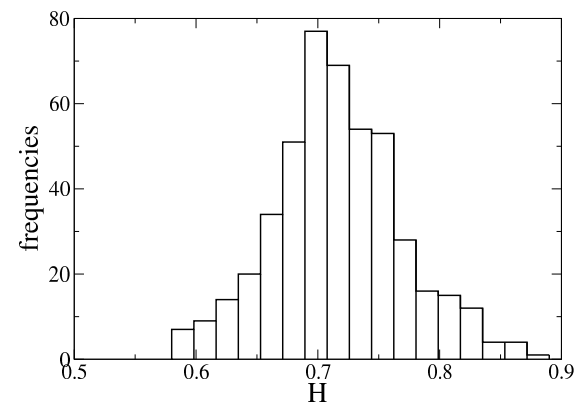}
\includegraphics[scale=0.3]{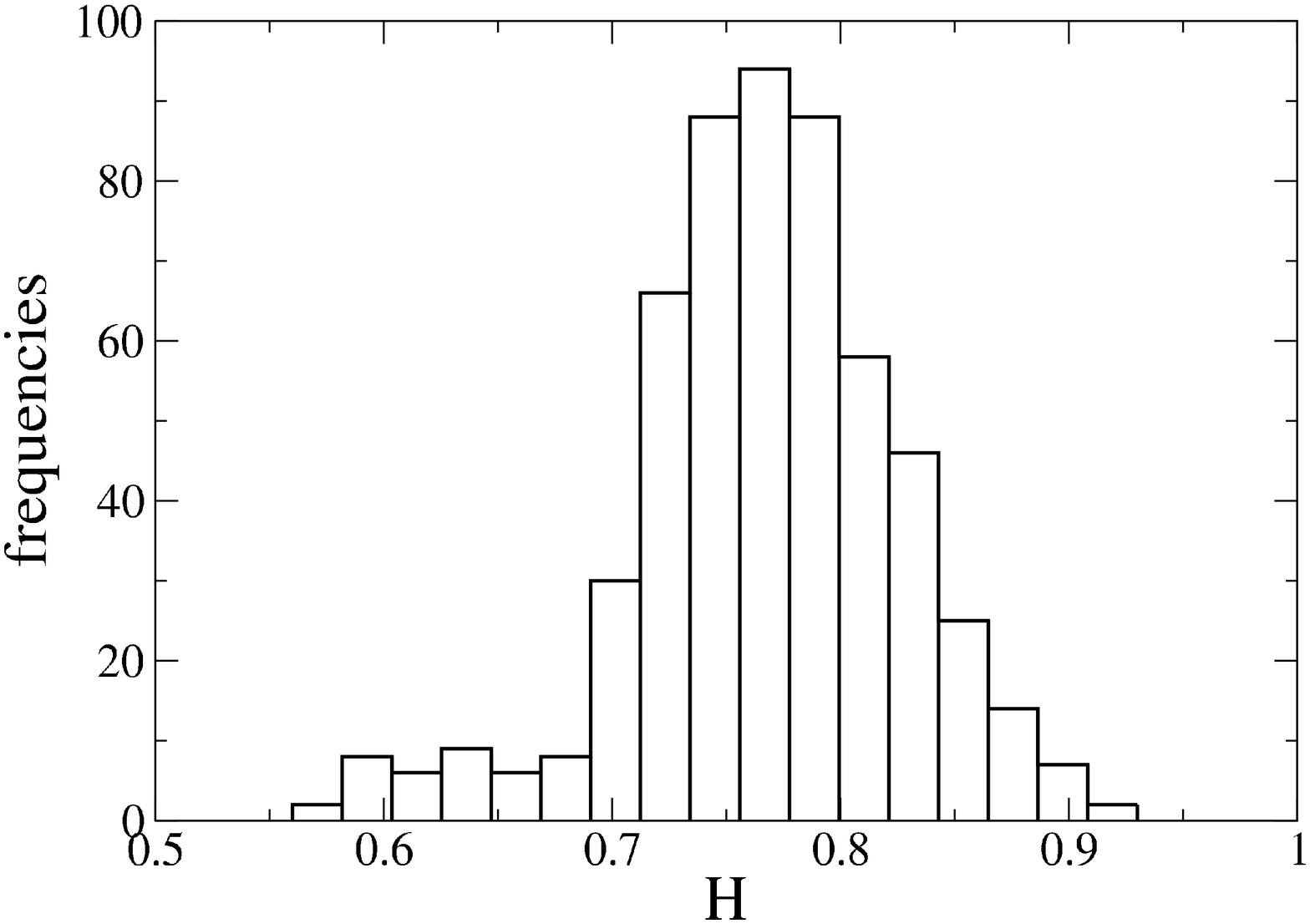}
\caption{Histogram of Hurst exponents for (a) market orders, (b) limit orders,
and (c) cancellations, for subsets of the data as described in the
text.  In (a) we show histograms for both the periodogram (continuous
line) and the R/S method (dashed line),
while (b) and (c) show the periodogram method
only.  The inset of (a) plots the results from the two methods against
each other (periodogram on the $x$-axis and R/S on the $y$ axis).}
\label{Hhist}
\end{center}
\end{figure}

The first estimator we used for the determination of the Hurst
exponent is least squares fitting of the periodogram. The mean
estimated value of the Hurst exponent is $H=0.695\pm0.039$ for market
orders, $H=0.716 \pm 0.054$ for limit orders, and $H=0.768 \pm 0.059$
for cancellations, where the error is the standard deviation. The
histograms of the exponents obtained in this way are shown in
Fig. \ref{Hhist}.
We see that in every case the Hurst exponent is roughly peaked around
the value $H=0.7$ which corresponds to $\alpha=0.6$.
 
Following the suggestion of Taqqu, Teverovsky and Willinger (1995) we also estimate the Hurst
exponent for market orders through the classical R/S method. 
In this case the mean Hurst exponent is
$0.696\pm 0.032$, which is consistent with the value obtained with the
periodogram method. Figure~\ref{Hhist}(a) gives a comparison of the
results of the two methods.  In the inset we plot the Hurst exponent
obtained from the periodogram against the Hurst exponent obtained from
the R/S method, showing that the results are quite correlated on a
case-by-case basis, with no discernable bias.

\subsection{Idiosyncratic variation of the Hurst exponents}

The previous results bring up the interesting question of whether
there are real variations in the Hurst exponents, or whether they have
a universal value, and the variations that we see are just sample
fluctuations.  To compare the longtitudinal and cross-sectional
variations we perform a classical ANOVA test. We assume that for each
stock $i$ the value of the Hurst exponent in different time periods is
normally distributed with mean $m_i$ and standard deviation
$\sigma$. We test the null hypothesis that all the $m_i$ are equal. We
indicate with $H_{ij}$ the estimated Hurst exponent of stock $i$ in
sub-period $j$. There are $r=20$ stocks, each of them with a variable
number $n_i$ of sub-periods. The total number of subsets is $n=\sum_i
n_i$.  As usual the sum of squares of deviations of $H_{ij}$ can be
decomposed in the sum of squared deviations within groups
(i.e. stocks) $(n-r)s_2^2 = \sum_{i=1}^r \sum_{j=1}^{n_i} (H_{ij} -
\bar{H}_i)^2$ and the sum of squared deviations between groups
$(r-1)s_1^2 = \sum_{i=1}^r (\bar{H}_i - \bar{H})^2$, where $\bar{H}$
is the sample mean for the entire sample and $\bar{H}_i$ is the sample mean
for stock $i$.  Under the above null hypothesis, the sum of squared
deviations within groups has a $\chi^2$ distribution with $n-r$ degrees
of freedom. Likewise the sum of squared deviations between groups has
a $\chi^2$ distribution with $r-1$ degrees of freedom. Therefore the
logarithm of the ratio between $s_1$ and $s_2$ has Fisher's
Z-distribution with $(r-1,n-r)$ degrees of freedom.  For all the three
types of orders we reject the null hypothesis with $99\%$
confidence. Moreover in all three cases $s_1>s_2$, showing that the
cross sectional variation of the Hurst exponent is significantly
larger that the longitudinal variation, which suggests that the
variations in the exponents between stocks are statistically
significant.  Nonetheless, it is interesting that these variations are
relatively small.

\subsection{Order flow in real time \label{realTime}}

We have shown that the sequence of order signs is a long-memory
process in event time. In this section we briefly consider the
correlation properties in real time. This is complicated by the fact
that trading is not homogeneous.  There are both strong intra-day
periodicities, e.g. volume tends to increase near the open and the
close, and also strong temporal autocorrelations in the number of
trades.  Thus the number of trades in any given time interval of
length $T$ can vary dramatically.

\begin{figure}[ptb]
\begin{center}
\includegraphics[scale=0.3]{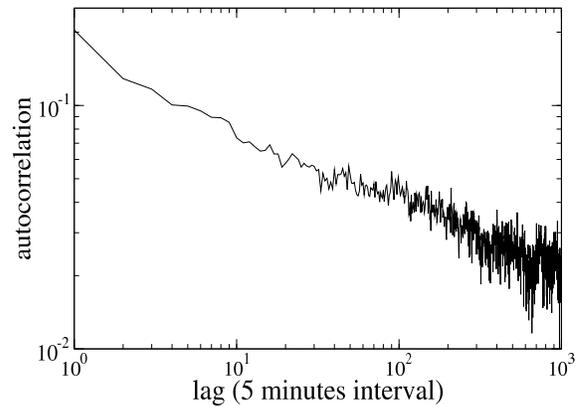}
\caption{
Autocorrelation function of the quantity 
$(n_b(t)-n_s(t))/(n_b(t)+n_s(t))$, where $n_b(t)$ and $n_s(t)$ indicate
the number of buy and sell market orders, respectively,
in a time interval of length $T=5$ minutes starting 
at time $t$. A lag unit on the $x$ axis corresponds to $5$
minutes.  There are strong positive autocorrelations for periods
of at least $5000$ minutes, or roughly $10$ days, and no indication
that there is anything special about the daily boundary.}
\label{time}
\end{center}
\end{figure}

To understand the long-memory of orders in real time, we are seeking a
quantity that gives information about imbalances in the signs of
orders, but which is independent of the number of orders placed in a
given time interval.  We use two methods, which give similar results.
Let $n_b(t)$ and $n_s(t)$ indicate the number of buy and sell market
orders, respectively, in a time interval of length $T$ starting at
time $t$. The first method follows a
majority rule, which assigns the value $+1$
if $n_b(t)>n_s(t)$ and the value $-1$ if $n_b(t)<n_s(t)$.  When
$n_b(t)=n_s(t)$ or there are no market orders in the interval we
assign the value $0$. The main defect of this method is that it
does not distinguish intervals with small or large imbalance of one type
of orders.  The second method is to use a continuous variable defined as
$(n_b(t)-n_s(t))/(n_b(t)+n_s(t))$ when $n_b(t)+n_s(t)\ne 0$
and zero elsewhere. This is bounded between $-1$ and
$1$. In Figure~\ref{time} we show the autocorrelation function of
$(n_b(t)-n_s(t))/(n_b(t)+n_s(t))$ for a time interval $T=5$ minutes for Vodafone. We note that a
power law decay of the autocorrelation function fits the
empirical data quite well, with an exponent $\alpha=0.3$, which is close to the
corresponding value in event time. 

This study makes it quite clear that the long-memory properties of
order signs persist across trading days.  There are $102$ intervals of
length $5$ minutes in a trading day, which means that the last lag in
Figure~\ref{time} corresponds to approximately $10$ trading days. 
Moreover the
autocorrelation does not show any significant peak or break in slope
near lag = $102$, indicating that the long-memory properties of the
market persist more or less unchanged across daily boundaries.

\subsection{Autocorrelation of transaction volume}

\begin{figure}[ptb]
\begin{center}
\includegraphics[scale=0.3]{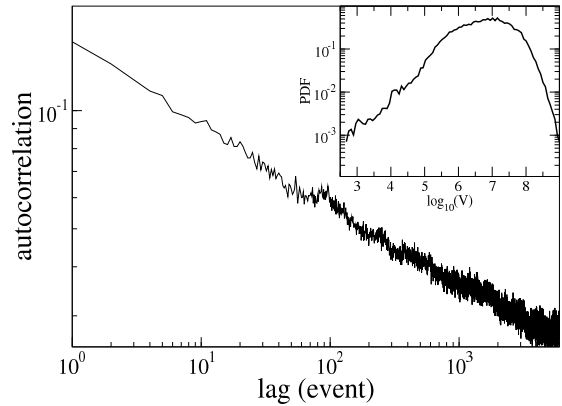}
\caption{
Autocorrelation function of the sequence of transaction values
for Vodafone in the period 1999-2002. The transaction value is
defined as the product of the number of shares times the price.
The lag is measured in terms of the number of transactions.
The inset shows the unconditional histogram of transaction value in
double logarithmic scale.}
\label{volautoc}
\end{center}
\end{figure}

We now show that the volume of the transactions is a long
memory process in event time; later on in Section~\ref{anticor} we
will argue that this is connected to the long-memory properties of order
signs via market efficiency.  The long-memory
properties of aggregated volume have been known for a long time
(Lobato and Velasco, 2000, Gopikrishnan et al., 2000).
 We use modified R/S statistics in order
to test the null hypothesis that the transaction volume is a short
memory process in event time. The value of a stock changes in time
because of the change in price. Therefore one could expect that the
number of traded shares is non-stationary due to the non stationarity
of the price. For this reason we decide to investigate the {\it value}
of the transaction, defined as the product of the number of traded
shares and the transaction price. The value is invariant under stock
splits.  In Figure~\ref{volautoc} we show the autocorrelation function
of the volume of Vodafone measured in terms of value in the period 1999-2002.
The inset shows a histogram of the transaction volume, which is well
fit by a Gamma distribution.  Once we adjust for scale, this seems to
be roughly the same for all the stocks in the sample (see also Farmer and Lillo, 2004). 
Moreover this
result is in contrast with what has been observed for the NYSE
by Gopikrishnan et al. (2000).

We applied the modified R/S test to the $324$ subsets used to test the
long-memory properties of market order size.  One of the conditions
for the applicability of the modified R/S test is that the
unconditional kurtosis of the time series is finite; from the
inset of Figure \ref{volautoc} it seems that the volume distribution
does not have a power-law tail, so the modified R/S test is
applicable.  However there could be biases in the test due to large
fluctuations in volume.  In $303$ of the $324$ subsets,
or $94\%$ of the time, we reject the null hypothesis of short-memory
for the transaction volume with $95\%$ confidence.  The Hurst exponent
estimated with the periodogram method varies across subsets and the
mean value is $H=0.732\pm0.075$, within the sampling error of the exponent
found for order signs.

\section{Market inefficiency?\label{efficiency}}

At first sight the long-memory property of the market order sign
time series is puzzling when considered from the perspective of market
efficiency.  Long-memory implies strong predictability using a simple
linear model.  When this is combined with the fact that orders have
price impact, it naively suggests that price changes should follow a
long-memory process as well.  That is, buy market orders tend to drive
the price up, and sell market orders tend to drive it down. Thus, all
other things being equal, a run of buy orders should imply future
upward price movement, and a run of sell orders should imply future
downward price movement. The predictability of order signs is sufficiently
strong that one would expect that profits could be made by taking
advantage of it.

There are many ways to define market efficiency, and we should be
clear how we are using this term.  Here we mean specifically {\it
  linear efficiency}, i.e. that the series of price returns contains
negligible temporal autocorrelations.  This allows for the possibility
that there might be other more complicated nonlinear patterns, and assumes a
trivial reference equilibrium that supports an IID random price
process.  This is a strong notion of efficiency in the sense of Fama
(1970), in that the information set is a sequence of recent buy or
sell order signs, which for the LSE is publicly available during this
period in real time.  We will make a subjective judgement concerning
what we mean by ``negligible'', without making detailed estimates of
transaction costs: If the directional movements in price returns have
long-memory, then the market is unlikely to be efficient, whereas if
they have short-memory, it becomes much more difficult to tell.

In this section we explore the consequences of long-memory in order
signs, and show that its impact on the predictability for prices is
offset by other factors.  In particular, the relative size of buy and
sell market orders and the relative size of the best quotes at the
best bid and ask move in a way that is anti-correlated with the long
memory of order signs, and compensates to make the market more
linearly efficient.  While order signs, market order volume, and
volume at the best prices are all long-memory processes, directional
price changes do not appear to have this property.

\subsection{Inefficiency of prices in absence of liquidity fluctuations}

\begin{figure}[ptb]
\begin{center}
\includegraphics[scale=0.3]{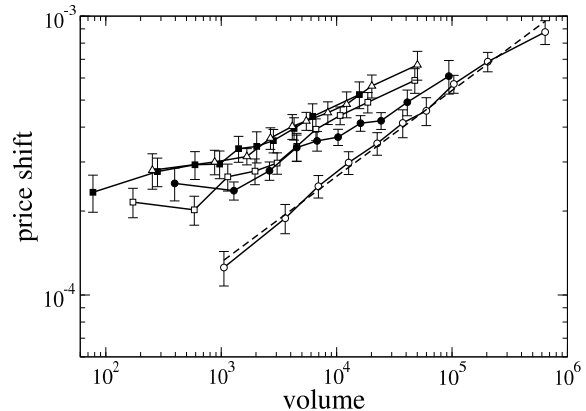}
\caption{
Market impact function of buy market orders for a set of 5 highly capitalized stocks traded
in the LSE, specifically AZN (filled squares), DGE (empty squares), LLOY
(triangles), SHEL (filled circles), and VOD (empty circles). Trades of
different sizes are binned together, and the average size of the
logarithmic price change for each bin is shown on the vertical
axis. The dashed line is the best fit of the market impact of VOD with
a functional form described in Eq.(9).  The value of the fitted
exponent for VOD is $\beta=0.3$.}
\label{impact}
\end{center}
\end{figure}

In this subsection we show that if liquidity were fixed, the
long-memory in the signs of orders would drive a strong inefficiency
in prices.  We first construct a series of surrogate prices assuming
that the reponse of prices to new market orders depends on order size
but is otherwise fixed.  The relation between the volume of a market
order and the consequent price shift is described by the average {\it
  price impact} function (also called the average {\it market impact}
function).  Recent studies of the impact of a single transaction
(Hasbrouck, 1991, Hausman and Lo, 1992, Farmer, 1996, Potters and
Bouchaud, 2002, Lillo, Farmer and Mantegna, 2003) have shown that the
average market impact is a concave function of either order or
transaction volume, matching other studies based on time-aggregated
volume (Torre, 1997, Kempf and Korn, 1999, Plerou et al., 2001, Evans
and Lyons, 2002).  It appears that the average impact varies across
markets and stocks. For example, for a set of $1000$ stocks traded at
NYSE (which works with a specialist) the impact is roughly
\begin{equation}
E(r|V)=\frac{sign(V) |V|^{\beta}}{\lambda}
\label{priceImpact}
\end{equation}
where $r$ is the logarithmic price return, $V$ is the volume of a
transaction, and $\lambda$ is a liquidity parameter.  The exponent
$\beta$ depends on $V$ and is approximately $0.5$ for small volumes
and $0.2$ for large volumes (Lillo, Farmer and Mantegna, 2003). The
liquidity parameter $\lambda$ varies for each stock, and in general
may also vary in time.  Potters and Bouchaud (2002) analyzed a much
smaller set of stocks traded at the Paris Bourse and NASDAQ and
suggested a logarithmic price impact function.  For the LSE,
Figure~\ref{impact} shows the price impact of buy market orders for 5
highly capitalized stocks, i.e. AZN, DGE, LLOY, SHEL, and VOD.  
The
price impact is well fit by the relation $E(r| V) \propto V^{\beta}$,
where $V$ should now be interpreted as the market order size $V = |
\omega |$ and $\beta\simeq 0.3$.

\begin{figure}[ptb]
\begin{center}
\includegraphics[scale=0.3]{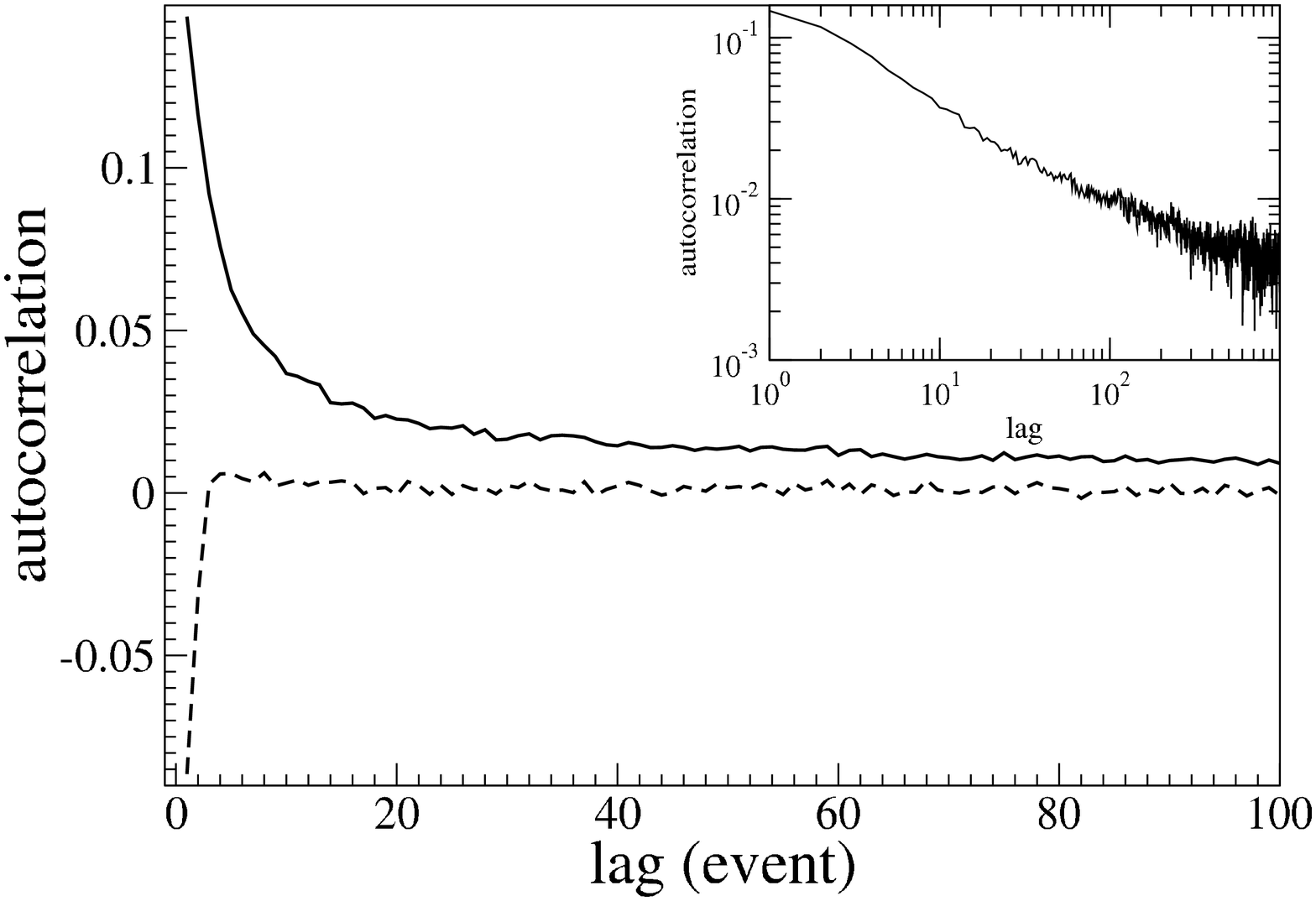}
\caption{Autocorrelation function of two surrogate time series of prices. 
The continuous line is the autocorrelation function of the surrogate time series
obtained by using the real order flow of market order of Vodafone
(volume and sign) and by using a deterministic price impact of Eq.(9)
with $\beta=0.3$. The dashed line is the autocorrelation function of the 
surrogate time series obtained by the real price shift due only
to market orders for Vodafone. In the inset we show the autocorrelation
function of the surrogate time series obtained with the deterministic price impact 
in a double logarithmic scale.}
\label{diffusion_autoc}
\end{center}
\end{figure}

If one assumes that the price impact is a deterministic function of
order size, since market order placement constitutes a long-memory
process, the generated price return time series will be long-memory
too. We test this conclusion by constructing a synthetic price time
series using real market order flow with a deterministic impact
function of the form of Equation~(9), but with $V$ now representing
market order size.  We use $\beta = 0.3$ as measured for Vodafone in
Figure~\ref{impact}, and arbitrarily set $\lambda = 1$. For each real
market order of volume $V_i$ and sign $\epsilon_i=\pm1$ we construct
the surrogate price shift $\Delta p_i=\epsilon_i V_i^{0.3}$.
In Figure~\ref{diffusion_autoc} we plot the autocorrelation function
of the surrogate time series of price shift obtained with the deterministic
price impact (upper curve). The inset of Figure~\ref{diffusion_autoc}
shows the same data with a double logarithmic scale. As expected the 
autocorrelation decays as a power-law, implying that
synthetic price returns are described by a long-memory process, in
contradiction with the assumption of linear efficiency. The application
of the Lo test rejects the null hypothesis of short memory and 
 the periodogram method gives a value of the Hurst exponent $H=0.66$.


There are two possible explanations for how the real price series can
be efficient when the surrogate price series defined above are
inefficient. We have only used market orders above, so the first
possible reason is that the price shift generated by limit orders and
cancellations act to make the market efficient.  The second
possibility is that the assumption of deterministic price impact is
wrong and efficiency comes about due to fluctuations in the impact.
The first reason has been recently suggested by Bouchaud et
al. (2004).  Their argument is that the price shift due to a market
order is anticorrelated with the price shift generated by limit orders
and cancellations placed between market orders. We verified
empirically that such an anticorrelation does exist.  However, as we
will show below, the market is approximately efficient, in the sense
that directional price changes do not display long-memory, even when
we include only price shifts driven by market orders.  Instead, we
show that efficiency is due to fluctuations in liquidity.

To show that efficiency does not depend on limit orders and
cancellations, we plot in Figure~\ref{diffusion_autoc} the
autocorrelation function of the real price shift time series $\Delta
p_i$ due to each market order $i$.  We note that the sample
first-order autocorrelation coefficient $\hat \rho$ is negative. For
values of the lag between $2$ and $\simeq 10$ the autocorrelation is
positive and for lags larger than $10$ it fluctuates around zero. This
is in strong contrast to the autocorrelation function of the order
flow, which is consistently positive for at least the first $10,000$
lags.  This makes it clear that the long-memory of the order series
has been strongly suppressed, and suggests that it no longer exists.

In an attempt to test this more carefully we apply the Lo test.
However, this presents two main problems. First, the distribution of
price shifts due to market orders seems to have a power-law tail with
a relatively small tail index (Farmer and Lillo, 2004).  Therefore we
are not sure that the fourth moment of the price shift is finite as
needed for the statistics used in the Lo test to be valid.  Second,
$\hat \rho$ is negative and we cannot use Eq. (5) to find the optimal
value of $q$.  Despite these difficulties we applied the Lo's test to
the time series of price shifts due to market orders but we find
inconclusive results. Since we do not have an optimal value for $q$,
we calculated $Q_n(q)$ for different values of $q$ and we found that
for some values of $q$, $Q_n(q)$ is outside the $95\%$ confidence
interval for the null hypothesis of short memory, whereas for other
values is inside this confidence interval.  A clearer result is
obtained by computing the Hurst exponent of the time series with the
periodogram method. The value $H=0.53$ is very close to the value
$0.5$ expected for a short-memory process. In conclusion the surrogate
time series obtained with the deterministic price impact is clearly
long-memory, whereas the time series of market order-driven price
shifts is significantly less correlated with a strong suggestion that
it is probably a short memory process. Similar results are seen when
we compute the Hurst exponent for returns generated by limit orders or
cancellations alone, and for other stocks, see
Table~\ref{SeparatedReturns}.
\begin{table}
\caption{Hurst exponents for the four stocks Astrazeneca, Lloyds,
  Shell, and Vodafone, separating returns triggered by market orders,
  limit orders, and cancellations.}
\begin{tabular}{r|ccc|c}
stock & market orders & limit orders & cancellations & all events\\
\tableline
AZN & 0.54 & 0.54 & 0.53 & 0.47\\
LLOY & 0.54 & 0.54 & 0.51 & 0.50\\
SHEL & 0.52 & 0.50 & 0.43 & 0.41 \\
VOD & 0.54 & 0.50 & 0.51 & 0.50\\
\end{tabular}
\label{SeparatedReturns}
\end{table}
This shows that market order, limit order, or cancellation-driven
fluctuations are approximately efficient when considered by
themselves\footnote{There is a slight tendency for the Hurst exponent
  of all events to be a little smaller than the Hurst exponent of each
  type of event taken separately.  This suggests a cooperative effect
  in which one type of event mean reverts against the other, as
  suggested by Bouchaud et al. (2004).}.

\subsection{The key role of fluctuations in relative
liquidity \label{anticor}}

In this section we will study fluctuations in liquidity and
fluctuations in market order size, and show that the ratio of the two
responds to changes in order sign predictability so as to make the
market more efficient.

It is clear that the liquidity, e.g. defined as the $\lambda$
parameter in equation~\ref{priceImpact}, makes large variations in
time.  For most purposes it is not a good approximation to treat it as
a constant.  A study of the LSE makes it quite clear that
fluctuations in liquidity are large in comparison to the volume
dependence of $E[\lambda_i | V]$, and that in
Equation~\ref{priceImpact} one should regard $\lambda$ as a random
variable whose fluctuations are roughly as large in relative terms as
those of $\Delta p$ (Farmer et al., 2004).

We first show that uncorrelated fluctuations in liquidity are not
sufficient to ensure linear efficiency, and then return to study the
correlations.  Consider Equation~\ref{priceImpact} and let us assume
that the inverse of the liquidity $\ell_i\equiv 1/\lambda_i$ is a
random variable uncorrelated with market order sign and size. In the
previous section we have seen that $a_i\equiv\epsilon_i V_i^{\beta}$
is a long-memory random process. Therefore if $E(a_i)=0$ then
$E(\Delta p_i)=0$, and the auto-covariance of price return is
\begin{eqnarray}
\gamma_{\Delta p}(\tau)=E(\Delta p_{i+\tau} \Delta p_i)=\nonumber\\
E(a_{i+\tau} a_i \ell_{i+\tau} \ell_i)=
E(a_{i+\tau} a_i)E(\ell_{i+\tau} \ell_i)=\nonumber\\
\gamma_{a}(\tau)(\gamma_{\ell}(\tau)+E(\ell)^2)
\end{eqnarray} 
Now $\ell$ is by definition a positive quantity and $E(\ell)> 0$.
Therefore the term in brackets in the last line of Eq. (10) cannot
be zero and $\gamma_{\Delta p}(\tau)\neq 0$, i.e. when the liquidity
is uncorrelated with the order flow, the market cannot be efficient. 

We will define the term {\it relative liquidity} to mean variations in
what is offered by liquidity providers, relative to what is being
asked for by liquidity takers.  For our purposes here, we will define
this more precisely as the size of market orders relative to the
volume at the opposite best price (e.g., the size of a buy market
order relative to the volume at the best ask).

Our working hypothesis is that market efficiency is strongly
influenced by relative liquidity.  This hypothesis is influenced by
other work (Farmer et. al, 2004), in which we demonstrate the
following microscopic picture of market impact.  There we show that
market impact varies, both because the depth of stored limit orders
varies, and because the size of market orders varies with it.  Market
order placement and the volume (depth) at the best price are highly
correlated.  It is very rare for a market order to be larger than the
depth at the opposite best price (Farmer et al., 2004), presumably
because liquidity takers are reluctant to execute at prices worse than
the best price.  When market orders do trigger price changes, they
almost always do so by exactly removing the volume at the (opposite)
best price.  The size of the resulting change in the best price is
just the size of the gap between the best price and the next price
occupied by a limit order.  Thus, the market impact depends on two
quantities: Whether or not a market order is big enough to cause a
price change at all, and when it does cause a price change, the size
of the gap to the next occupied price.

How can a predictable run of orders of a given sign be consistent with
market efficiency?  For example, consider a period in which there has
been a run of buy market orders.  The long-memory implies that the
next order is more likely to be a buy order.  Our hypothesis is that
this is compensated by the fact that the next buy market order is less
likely to penetrate the best price.  This can be either because the
volume of limit orders at the best ask is larger, or because the
volume of the next buy market order is smaller, or both of the above.

In order to test this hypothesis we explicitly construct a model to
predict the sign $\epsilon_t$ of the next order.  We do this by making
an autoregressive model of the form 
\begin{equation}
\hat{\epsilon}_{t} = \sum_{i=1}^{N} a_i \epsilon_{t-i}.
\label{SignPredictor}
\end{equation}
To fit the coefficients of the model we use ordinary least squares.
The values of the largest lag $N$ vary depending on the stock, but are
typically the order of fifty.  $\hat{\epsilon}_t \simeq 0$ corresponds
to low predictability, whereas large values of $|\hat{\epsilon}_t|$
correspond to high predictability.  We have tested this and
demonstrated that it works very well.  When $\hat{\epsilon}_t$ is
$0.5$, for example, there is a $75\%$ probability that the next order
is a buy order\footnote{A more proper approach would be to use a model
  specifically tailored for forecasting probabilities, but the quick
  and dirty approach above is sufficient for our purposes here.}.

\begin{figure}[ptb]
\begin{center}
\includegraphics[scale=0.3]{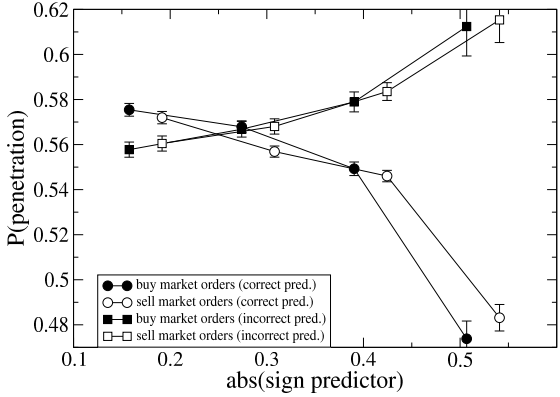}
\caption{The probability that the next order penetrates the price,
  conditioned on the value of the sign predictor of
  equation~\ref{SignPredictor}.  The cases where the sign prediction
  is correct are shown as circles, and where it is incorrect as
  squares.  Buys are solid and sells are empty.  The bins correspond
  to equal intervals in $\hat{\epsilon_t}$.  The error bars shown are
  standard errors, which are clearly too optimistic given the
  long-memory of the data. These results are shown for AZN; we see
  similar results for other stocks.}
\label{penetration}
\end{center}
\end{figure}

We now test our hypothesis by computing the probability of a
penetration of the best price as a function of the strength of the
sign predictor.  (We say that a market order {\it penetrates} when it
is as large or larger than the opposite best, and so causes a
mid-price change).  We analyze this for buying and selling, and for
predictions that are right or wrong.  For example, when the sign
predictor is positive, predicting that the next order will be a buy
order, we compute the fraction of buy market orders that penetrate the
best ask.  We compare this to the case when the sign prediction is
also positive, but the prediction is wrong, in which case we compute
the fraction of sell orders penetrating the best bid.  We repeat all
this similarly with signs reversed when the sign predictor is
negative.  We bin based on the value of the sign predictor and compute
the probabilities for each bin.  The results are shown in
Figure~\ref{penetration}.
We see a large variation in the results.  When the sign is
unpredictable, the predictor has a value near zero.  In this case the
penetration probability is roughly $58\%$ when the sign predictor is
correct, and about $56\%$ when it is incorrect (for both buys and
sells).  But when the predictability is high, i.e. when
$|\hat{\epsilon}_t| \simeq 0.5$, the situation is quite different.
When the prediction is correct, the probability of penetration is much
lower, roughly $48\%$, and when it is incorrect it is much higher,
about $61\%$.  This shows that when predictability is high the market
acts to decrease the probability that an order of the predicted sign
will penetrate.  This can be achieved either through higher volume at
the opposing best, or by smaller market order size, or both.  In
either case, this is what we mean when we say that the relative
liquidity acts to oppose the trend in order flow.  We see similar
results for other stocks\footnote{The level of the penetration
  probabilities shifts from stock to stock, but the basic pattern
  remains the same.}.

\begin{figure}[ptb]
\begin{center}
\includegraphics[scale=0.3]{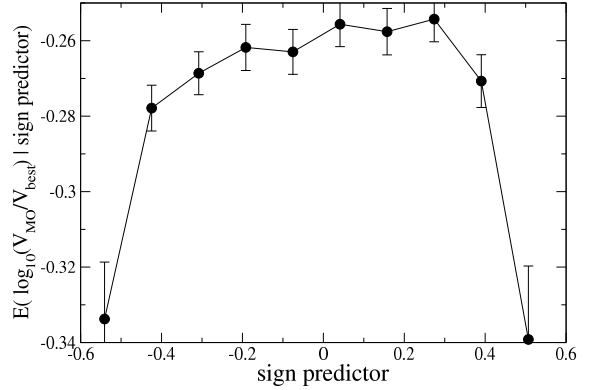}
\caption{The expected value of the logarithm of the ratio of market
  order volume to volume at the opposite best price, conditioned on
  the value of the sign predictor $\hat{\epsilon_t}$ of
  equation~\ref{SignPredictor}.  The bins correspond to ten equal
  intervals in $\hat{\epsilon_t}$; the expectation is the average
  value in each bin. The error bars shown are standard errors, which
  are clearly too optimistic given the long-memory of the data. These
  results are for the stock AZN; we see similar results for other
  stocks.}
\label{volumeratio}
\end{center}
\end{figure}

To test this hypothesis in a different way we plot the expected value
of the logarithm of the ratio of the market order size to the volume
at the opposite best, as a function of the strength of the order sign
predictor of equation~\ref{SignPredictor}.  This includes all events,
whether the sign prediction is right or wrong.  The results for the
stock Astrazeneca are shown in Figure~\ref{volumeratio}.
Corresponding to our previous result, we see that the ratio of market
order size to volume at best is larger when the sign is unpredictable,
and smaller when it is predictable.  This reinforces our conclusion
that fluctuations in relative liquidity oppose trends in order signs.
Similar results are observed for other stocks.

An obvious question is whether these results are primarily driven by
changes in market order size, or by changes in the volume at the best.
We have done a variety of tests for this, with inconclusive results.
We have seen some indications that variation in volume at the best
conditioned to the sign predictor is the dominant effect, but the
effect is only $1.5\%$, in contrast to the effects above, which are on
the order of $20\%$.  The problem is that there are large covarying
level shifts in volume and liquidity through time.  Their ratio, in
contrast, is not affected by such level shifts, and is a much more
sensitive indicator.

The other question one naturally asks concerns the magnitude of price
responses.  Given that a market order penetrates the best price, are
there significant variations in the size of the resulting price
response?  In other words, are there asymmetries in the gaps in the
two sides of the limit order book that are conditioned on the sign
predictor?  Preliminary studies suggest that this is not as important
as the change in the probability of penetration.

\begin{figure}[ptb]
\begin{center}
\includegraphics[scale=0.3]{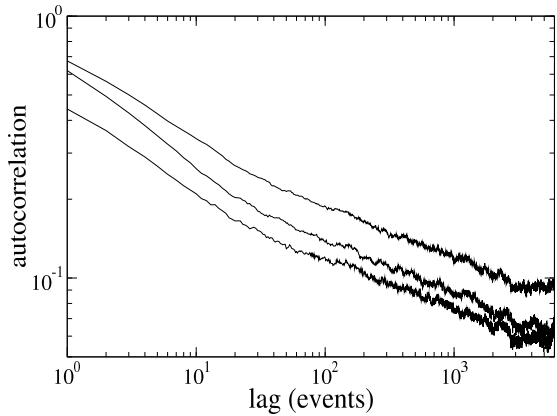}
\caption{The autocorrelation of the volume at the best prices, shown
in double logarithmic scale, as a function of time measured in terms
of the number of market orders. The three curves shown, from
top-to-bottom, are the volume at the best ask, best bid, and best price
(i.e. the best ask when the order is a buy order and the best bid when
the order is a buy order).  All three are long-memory processes.}
\label{liquidityMem}
\end{center}
\end{figure}

These results suggest that the volume of market orders and the volume
at the best price are comoving with trends in order signs in order to
make the market more efficient.  This motivated us to test whether or
not these are long-memory processes.  Typical autocorrelation
functions are shown in Figure~\ref{liquidityMem}.  
These resemble the autocorrelation functions for order signs, making
it quite clear that they are also long-memory processes.  In
presenting the results in this order, we do not mean to necessarily
suggest that the long-memory of order signs is primary, and that the
long-memory of volume and liquidity are consequences of it, but rather
to say that these phenomena are intimately related: From the analyses
presented here, one could equally well say that the long-memory of
order signs adjusts in order to offset that of volume and liquidity.
The key point is that to enhance linear efficiency, despite
the long-memory of all three of these processes, they must be in a
certain sense out of phase, so that their effect on prices roughly
cancels.

We have not demonstrated that these effects are sufficient to ensure
linear market efficiency.  Rather we have demonstrated that they are
quite strong, and they act in the right direction to make the market
more efficient.  In the conclusions we discuss some possible motives for
this behavior.

\section{Individual institutions\label{investors}}

In this section we consider the behavior of individual institutions in
order to gain some understanding of what drives the long-memory
processes described above. The LSE database allows us to track the
actions of individual institutions through a numerical code which
identifies the institution. For privacy reasons the code is different
for different stocks and it is reshuffled each calendar
month. Therefore our analysis will be limited to a single trading
month.

We consider as a case study the market order placement of Vodafone in
July 2002. Our choice is motivated by the fact that Vodafone is one of
the most heavily traded stocks.  In this month there were $45,774$
market orders distributed across $155$ trading institutions. We have
found that the $12$ most active institutions are responsible for more
than $70\%$ of market orders.  Thus, the participation in
trading is extremely inhomogeneous among the institutions,
with a few institutions placing many orders and many
institutions placing only a few orders.
\begin{table}
\caption{Summary statistics of the $12$ most active institutions
trading Vodafone in July 2002.}
\begin{tabular}{rccc}
code & number of market orders & fraction of buy & $Q_n$\\
\tableline
3589& 6065& 0.24& {\bf 2.27}\\
2146& 3929& 0.53& 1.33\\
1666& 3606& 0.54& {\bf 1.99}\\
3664& 3532& 0.44& {\bf 2.85}\\
1556& 3291& 0.50& {\bf 1.88}\\
3007& 3132& 0.49& 1.06\\
1886& 2154& 0.51& {\bf 2.35}\\
1994& 1530& 0.32& {\bf 2.66}\\
2196& 1512& 0.51& 1.50\\
2681& 1420& 0.52& {\bf 1.99}\\
2742& 1247& 0.49& 1.66\\
 823& 1200& 0.41& 1.59\\
\end{tabular}
\label{active}
\end{table}

\begin{figure}[ptb]
\begin{center}
\includegraphics[scale=0.3]{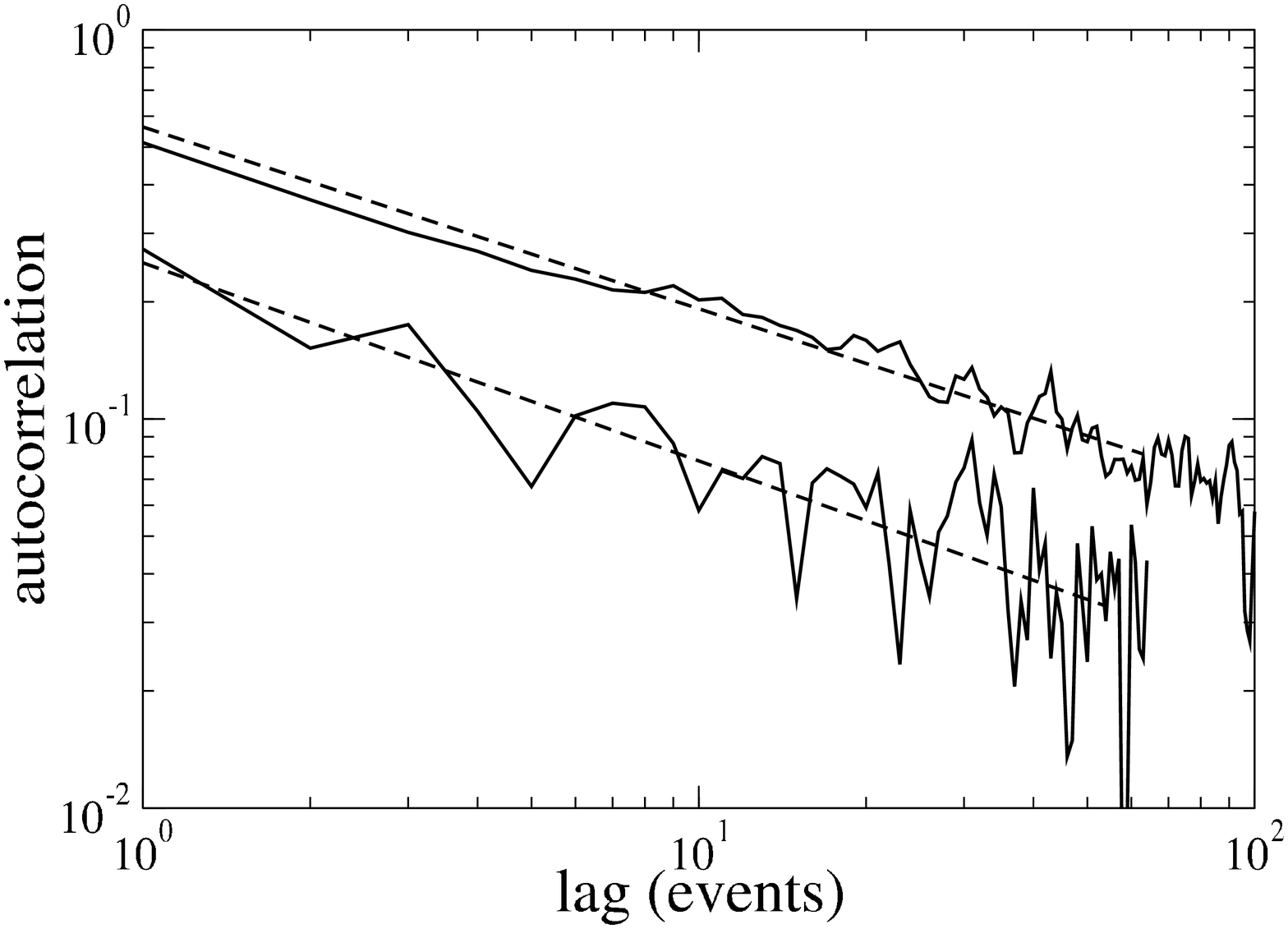}
\includegraphics[scale=0.3]{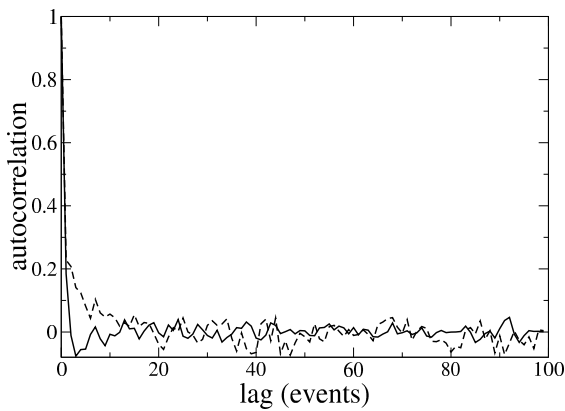}
\caption{Autocorrelation function of the market order sign time series
  for four institutions trading Vodafone in July 2002. In (a) we show
  two institutions displaying long-range memory in market order
  placement. The top curve refers to institution 3589 and the bottom
  one to institution 1886.  The dashed lines are the best fit of the
  empirical curves with a power law.  The exponents are $\alpha=0.47$
  and $\alpha=0.51$, respectively.  In (b) we show two institutions
  displaying short range order placement, institution 3007 (continuous
  line) and institution 823 (dashed line). In both panels the lag is
  measured as the number of market orders.}
\label{autoc-investors}
\end{center}
\end{figure}

Table~\ref{active}
shows the identification code, the number of market orders, and the
fraction of market orders that are buy orders for each of the twelve
largest institutions.  In Figure~\ref{autoc-investors} we show the
autocorrelation function of the time series of market order signs
for four active institutions.
In panel (a) we show two institutions whose market order flow is a
long-memory process. One of the two institutions (code 3589) is the most
active institution, which placed buy market orders $24\%$ of the time,
and the other one (code 1886) placed buy market orders $51\%$ of the
time.  We see that in both cases the autocorrelation function is
well-described by a power law with an exponent $\alpha\simeq 0.5$,
which corresponds to $H=0.75$. Panel (b) shows two active institutions
(code 3007 and code 823) whose market order sign time series is a
short-memory process.  To test the hypothesis that the individual
market order placement is a long-memory process more rigorously, we
apply the modified R/S test to the time series of the market order
signs of the twelve most active institutions. Table~\ref{active}
reports the value of $Q_n$.  A boldface font indicates the cases when
$Q_n$ is outside the $95\%$ confidence interval of the null hypothesis
of short-memory. We see that for $7$ of the $12$ active institutions
we reject the null hypothesis of short-memory process.

We repeated these results for the stock AZN in August 2001 and got
similar results.  Out of the top ten institutions, according to the Lo
test, five clearly displayed long-memory, and five did not.

This result shows that even at the institution level the placement of
orders has long-memory properties. This is not true for all
institutions, but rather there is an heterogeneity in their
behavior. A correlated sign in the order placement could be an
indication of splitting a large order in smaller size orders in order
to maximize profit without paying too much in terms of price
impact. On the other hand an uncorrelated (or at least short range
correlated) sign in the order placement could indicate different
strategies such as, for example, market making.  In section
\ref{causes} we suggest and discuss possible causes of the long-memory
of order flow.

\begin{figure}[ptb]
\begin{center}
\includegraphics[scale=0.3]{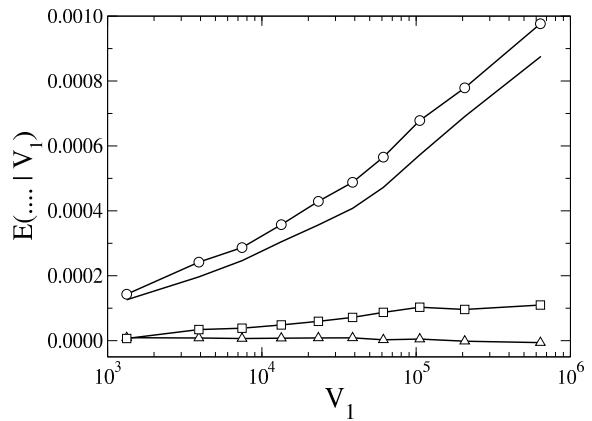}
\includegraphics[scale=0.3]{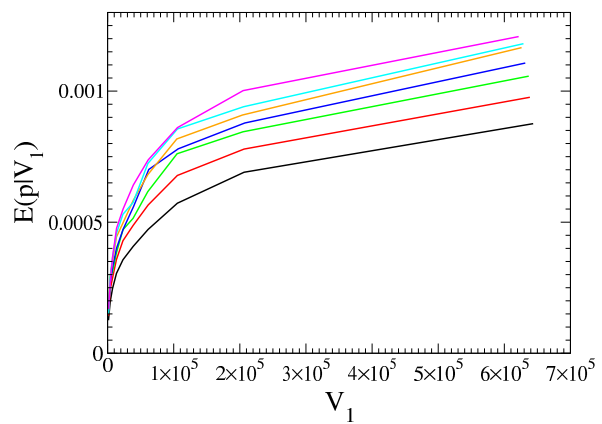}
\caption{ (a) A decomposition of the average delayed market impact of
  two successive market orders, as defined by Equation~\ref{decomposedImpact}, for
  Vodafone.  All four elements are conditioned on the size $V_1$ of
  the first market order.  The immediate impact of the first market
  order, $E(\Delta p_1|V_1)$ is the continuous line; the impact of any
  intervening limit orders or cancellations, $E(\Delta p_i|V_1)$, has
  triangles;, The immediate impact due to the second market order,
  $E(\Delta p_2|V_1)$, is shown with squares, and the total market
  impact $E(\Delta p_{1-2}|V_1)$ is shown with circles\\ (b) The
  average market impact for a series of market orders, conditioned on
  the volume $V_1$ of the first order, $E(\Delta p_{1-m}|V_1)$.  In
  ascending order in the plot, the curves are $m=1$ (black), $m=2$
  (red), $m=3$ (green), $m=4$ (blue), $m=5$ (orange) $m=6$ (cyan) and
  $m=10$ (magenta).  The average market impact builds steadily with
  each order; this is caused by the long-memory of the order sign and
  order size. both panels show the results for buy market orders}
\label{delayimpact}
\end{center}
\end{figure}

\section{Implications for market impact\label{implications}} 

In this section we discuss a practical consequence of long-memory of
order flow. We have seen in Section V.A that the market impact is a
concave function of volume. We may therefore ask about the price shift
in the future (in transaction time) given that an order of a given
volume and sign has arrived in the present. To be more specific, let
us consider a buy market order of volume $V_1$ occurring now. The
generated price shift $\Delta p_1$ is the difference between the
midprice just after the order and just before the order. Between this
market order and the next market order the midprice can change because
of new limit orders and cancellations, generating a price change
$\Delta p_i$. When the next market order arrives a new midprice shift
$\Delta p_2$ occurs.  The total price shift between the instant just
before the first order is placed and the instant just after the second
market order is placed is therefore
\begin{equation}
\Delta p_{1-2}=\Delta p_1+\Delta p_i+\Delta p_2
\label{decomposedImpact}
\end{equation} 
If the order flow were random we would expect that 
$E(\Delta p_{1-2}|V_1)=E(\Delta p_1|V_1)$ since the volume
and the sign of the next orders is uncorrelated with the
corresponding quantities of the first order. In Figure~\ref{delayimpact} we
present a decomposition of the impact of two successive market orders
in the terms described above. In panel (a) 
of Figure \ref{delayimpact} we show the four quantities
$E(\Delta p_1|V_1)$ (the same quantity shown in Fig. 
\ref{impact}), $E(\Delta p_i|V_1)$, $E(\Delta p_2|V_1)$
and $E(\Delta p_{1-2}|V_1)$. We see that $E(\Delta p_i|V_1)$ 
is almost zero, meaning that the price shift due to 
limit orders and cancellations after a market order 
is relatively unimportant.
This result suggests that the role of price reversion
due to limit orders and cancellations between two
market orders is marginal in making the market efficient.
On the other hand $E(\Delta p_2|V_1)$ 
is clearly positive and increasing with $V_1$. This is due
to the strong temporal correlation in market order sign and 
size. In fact if the first market order is a buy market
order it is probable that the next market order is also a buy and
the volume of the second market order is correlated with the first
one. Therefore it is more probable that the price will move up due
to the arrival of the second order. Figure~\ref{delayimpact} shows
$E(\Delta p_{1-2}|V_1)$, which is simply the sum of the three
terms, as shown in Eq.~\ref{decomposedImpact}. The distance between $E(\Delta p_{1-2}|V_1)$
and $E(\Delta p_1|V_1)$ is a measure of the effect of the correlation
of order sign and size in the delayed price impact.

To extend this analysis to more orders, we study the delayed market
impact $E(\Delta p_{1-m}|V_1)$ where $m$ is the number of future
market orders. Panel (b) of Figure \ref{delayimpact} shows this
quantity as a function of $V_1$ for $m=2,3,4,5,6$ and $m=10$. We see
that for a fixed value of $V_1$, $E(\Delta p_{1-m}|V_1)$ is an
increasing function of $m$. This is clearly due to the long
correlation of market order sign and size. Eventually, for large
values of $m$, $E(\Delta p_{1-m}|V_1)$ becomes independent of $m$.

\section{Conclusions\label{conclusion}}

\subsection{Comparison to work of Bouchaud et al. (2004)}

Now that we have presented all our results, we can compare to the work
of Bouchaud et al (2004).  They independently discovered the same long-memory
effect we have reported here for market order flow in the Paris Stock
Exchange.  We have taken the analysis in a somewhat different
direction than they have, and offered a different interpretation.
First, to convince potential sceptics, we have gone to extensive
length to demonstrate at a very high level of statistical significance
that order signs are indeed a long-memory process.  This is separately
true for market orders, limit orders, and cancellations.  We have
explicitly constructed a time series forecasting model that shows the
high degree of predictability that goes along with this behavior --
the conditional probability of the sign of the next order is
frequently as high as $75\%$.  As required to ensure market
efficiency, this predictability is not present in price movements.  We
have hypothesized that this is at least in part due to anti-correlated
changes in relative liquidity, as defined by the ratio of market order
size to volume at the best opposite price.  This is large effect,
involving variations in relative liquidity of the order of $20\%$.
This does not prove that this is sufficient to ensure efficiency, but
it does suggest that time varying relative liquidity plays a major role.

Bouchaud et al. (2004) have offered a different explanation.  The key
difference concerns the way in which liquidity is treated.  They
assume a constant mean-reverting propagator for market impact.
Whereas we have studied the way in which liquidity varies in
opposition to order flow, their assumption of a constant propagator
amounts to assuming that the liquidity is not varying in any
correlated manner with trends in order flow.  Instead, they assume
that the market impact is time dependent in a way tha causes much of
it to disappear.  They have pointed out that the market impact does
not grow as fast as one would naively expect, and have presented
evidence that it often reverts on a long timescale, corresonding to a
few hours to a day.  They propose that the key factor ensuring
efficiency is anti-correlated limit order placement, i.e. that later
placement of limit orders undoes the potentially long-memory permanent
price changes that would otherwise be caused by market orders.

We have presented two pieces of evidence that seem to oppose this
point of view: First, we show that market order, limit order, and
cancellation driven price changes are not long-memory, even when they
are considered individually.  This seems to imply that correlated
behavior of limit orders is not the effect that cancels the long-memory of
order flow.  Second, in our study of market impact in
Section~\ref{implications}, we showed that the expected price shift in
the intervening time between two market orders due to limit orders and
cancellations is relatively small.  It may still be possible, however,
that these small effects accumulate to become important, as suggested
by Bouchaud et al. (2004).

At this stage it is quite possible that both of these effects coexist:
While we have showed that time-varying liquidity is a significant
effect, we have not presented any evidence that it is the only effect.
It may well be that there are diverse forces working to ensure market
efficiency.

\subsection{Possible causes of long-memory order flow\label{causes}}

We have shown that the sign of order flow, order size, and liquidity,
are all long-memory processes. What might cause this? In this section
we make a few speculations about the possible origin of long-memory in
these fundamental inputs to price formation.

One possible explanation for long-memory in order flow is that it
simply reflects news arrival. Good (or bad) news may be clustered in
time, driving the sign of order flow.  Such news could either be
external to the market, or it could be generated by factors internal
to the market.  If external it could be a property of the natural
world, a reflection of the environment that humans necessarily
interact with.  We know that the intensity of floods, hurricanes,
earthquakes, and natural disasters have a power law distribution, and
perhaps these are just symptoms of a ubiquitous property of the
natural world that is reflected in what we consider ``news".
Alternatively, this could be an internal property, due to human social
dynamics.  Such ``news" might be internally generated, e.g. due to
herding behavior (Cont and Bouchaud, 2000), or it might be caused by
inattention: Time lags in the response of investors to news arrival
can cause autocorrelations in order flow.  However, it is not clear
why this should have a power law distribution.

A different explanation is in terms of the execution of large orders,
which leads to order splitting.  It is well-known that institutions
with large orders frequently split them into small pieces (Chan and
Lakonishok, 1993 and 1995), spreading out the execution of the orders
over periods that can be many months long.  If such orders have a
power law distribution, and the time needed to fully execute an order
is proportional to the size of the order, then this might give rise to
power law autocorrelations in time.  The idea that order size would
have a power law distribution is not implausible given that many
related quantities, such as firm size (Axtell. 2001), wealth (Pareto,
1896) and mutual fund size (Gabaix et al, 2003), have power law
distributions.

We have tended to discuss the autocorrelation of order signs as though
this were a primary property, and the behavior of volume and
liquidity are consequences, which must exist in order to maintain
market efficiency.  An alternative is that this reasoning is reversed,
and that the autocorrelation of volume and liquidity are primary
properties, and that of order signs is a consequence. However, it
still remains to be determined why any of these should have such strong
temporal autocorrelations. At this stage we simply don't know what
causes this phenomenon, but it is clearly a remarkable aspect of human
economic activity that deserves more attention.  

Even if we take the existence of long-memory as a given, we have not
explained the strategic behavior that ensures efficiency.  While the
correlated long-memory behavior of order signs, volume, and liquidity
are necessary to maintain efficiency, why are market
participants motivated to make this happen?  What profit-seeking
or risk-avoiding motives drive market participants to place orders in
such a way that order size and liquidity are anti-correlated with runs
in order signs?

To illustrate the problem, we will propose and then critique a
possible explanation based on market makers' incentive to control
prices. During a run of buy market orders, which would normally tend
to drive the price up, market makers as a group are by definition
sellers, and their positions become more negative.  If this is
accompanied by a price rise they will tend to lose money.  To prevent
this they might intentionally manipulate liquidity to prevent or
reduce upward price movements, by supplying more liquidity at the ask
than the bid.  In a specialist system in which a single agent has a
monopoly on market making this might be a reasonable explanation.
However, in a competitive environment such as the LSE, where there are
many market makers, this is more difficult to explain: The individual
who takes the lead in controlling the price will experience the
largest adverse change in position.  We know that on average buy
orders do tend to drive the price up (there is on average market
impact), so this behavior should systematically result in
losses\footnote{The key question is whether the profits from taking
  the spread are sufficient to offset the adverse price movements
  associated with long memory.}.  This suggests that either the market
makers as a group collude to control prices, sharing the burden
between them, or that this is not the correct explanation for this
behavior.

An alternative hypothesis that we consider more plausible is that
liquidity providers take advantage of trends in order flow to acquire
a desired position.  So, for example, consider a liquidity provider
that wants to sell, either to unload an existing inventory or to take
a new tactical position.  When a buying trend develops, she takes
advantage of it by placing larger sell limit orders at or near the
ask.  From this point of view, one might reverse the usual
terminology, and think of market orders as providing liquidity for
limit orders by making them more likely to be executed. Under this
hypothesis, liquidity providers take advantage of an imbalance in
market order flow, thereby damping the response of the price.

One clue about the long-memory behavior of order flow is that we see
it for some institutions and not for others.  This might be caused by
differences in order splitting strategies.  However, at this stage
this is just speculation.

\acknowledgments We would like to thank the James S. McDonnell
Foundation for their Studying Complex Systems Research Award, Credit
Suisse First Boston, McKinsey Corporation, Bob Maxfield, and Bill
Miller for supporting this research. We would also like to thank
J-P. Bouchaud, Dick Foster,and Laszlo Gillemot for valuable
conversations and Marcus Daniels for technical support

\newpage

\newpage

\newpage



\newpage

\begin{thebibliography}{99}

\bibitem{firmSize} Axtell, R., 2001, Zipf distribution of U.S. firm
  size.  {\it Science} {\bf 293}, 1818-1820.

\bibitem{backus} Backus, D.K. and S.E. Zin, 1993, long-memory
  inflation uncertainty: Evidence from the term structure of interest
  rates.  {\it Journal of Money, Credit and Banking} {\bf 25}, 681-700.

\bibitem{baillie96}
Baillie, R.T., 1996,
long-memory processes and fractional integration in econometrics.
{\it Journal of Econometrics} {\bf 73}, 5-59.

\bibitem{baillie95} Baillie, R.T., C.-F. Chung, and M.A. Tieslau,  1995,
Analyzing inflation by the fractional integrated ARFIMA-GARCH model.
{\it Journal of Applied Econometrics} {\bf 11}, 23-40.

\bibitem{FIGARCH} Baillie, R.T., T. Bollerslev, and H.-O. Mikkelsen, 1996,
Fractionally integrated generalized autoregressive conditional
heteroskedasticity.
{\it Journal of Econometrics} {\bf 74}, 3-30.

\bibitem{beran} Beran, J., 1994,
{\it Statistics for Long-Memory Processes}.
Chapman \& Hall.

\bibitem{bollerslev} Bollerslev, T. and H.-O. Mikkelsen, 1996,
Modeling and pricing long-memory in stock market volatility.
{\it Journal of Econometrics} {\bf 73}, 151-184.

\bibitem{bouchaud} Bouchaud, J.-P., Y. Gefen, M. Potters, and M. Wyart , 2004,
Fluctuations and response in financial markets: the subtle nature 
of 'random' price changes, 
{\it Quantitative Finance} {\bf 4} 176-190.

\bibitem{breidt} Breidt, F.J., N. Crato, and P.J.F. de Lima, 1993,
Modeling long-memory stochastic volatility.
Working paper (Johns Hopkins University, Baltimore, MD).

\bibitem{campbell}
Campbell, Y.J., A.W. Lo, and  A.C. Mackinlay,  1997,
{\it The Econometrics of Financial Markets}, 
(Princeton University Press, Princeton).

\bibitem{Lakonshok} Chan, L.K.C.  and J. Lakonishok, 1993,
Institutional trades and intraday stock price behavior.
{\it Journal of Financial Economics} {\bf 33}, 173-199.

\bibitem{Lakonshok2} Chan, L.K.C. and J. Lakonishok, 1995,
The behavior of stock prices around institutional trades.
{\it Journal of Finance} {\bf 50}, 1147-1174.

\bibitem{Cont98} Cont, R. and J.-P. Bouchaud, 2000,
Herd behavior and aggregate fluctuations in financial markets.
{\it Macroeconomic dynamics}. {\bf 4}, 170-196.

\bibitem{diebold89} Diebold, F.X. and G.D. Rudebusch, 1989,
long-memory and persistence in aggregate output.
{\it Journal of Monetary Economics} {\bf 24}, 189-209.

\bibitem{volRef}  Ding, Z., C.W.J. Granger, and R. Engle, 1993,
A long-memory property of stock market returns and a new model.
{\it Journal of Empirical Finance} {\bf 1}, 83-106.

\bibitem{embrechts} Embrechts, P., C. Kl\"uppelberg, and T. Mikosch, 1997,
{\it Modelling Extremal Events for Insurance and Finance}.
Springer-Verlag, Berlin Heidelberg.

\bibitem{Evans02} Evans, M.D. and Lyons, R.K., Order flow and exchange
  rate dynamics, {\it J. Political Economy} {\bf 110}, 170-180 (2002).

\bibitem {Fama70} Fama, E., 1970,
 Efficient capital markets: A review of theory and empirical work.
 {\it Journal of Finance} {\bf 25}, 383 - 417. 

\bibitem{Farmer96} Farmer, J.D., 1996,
 ``Slippage 1996'', 
Prediction Company internal technical report, http://www.predict.com

\bibitem{lilloQF} Farmer, J.D. and F. Lillo, 2004,
On the origin of power-law tails in price fluctuations.
{\it Quantitative Finance} {\bf 4}, C7-C11. 

\bibitem{mike} Farmer, J.D., Gillemot, L., Lillo, F., Mike, S., and
  Sen, A., 2004, What really causes large price changes?,
  in press in {\it quantitative Finance}. Preprint available at
  xxx.lanl.gov/cond-mat/0312703.

\bibitem{gabaix03} Gabaix, X., P. Gopikrishnan, V. Plerou, and
  H.E. Stanley, 2003, A theory for power-law distributions in
  financial market fluctuations. {\it Nature} {\bf 423}, 267-270.

\bibitem{stanleyvol} Gopikrishnan, P., V. Plerou, X. Gabaix, X., and H.E.
  Stanley, 2000, Statistical properties of share volume traded
  in financial markets.  {\it Physical Review E} {\bf 62},
  R4493-R4496.

\bibitem{granger} Granger, C.W.J. and R. Joyeux, 1980,
An introduction to long-range time series models and fractional differencing.
{\it Journal of Time Series Analysis} {\bf 1}, 15-30.

\bibitem{greene} Greene, M. and B. Fielitz, 1977,
Long-term dependence in common stock returns.
{\it Journal of Financial Economics} {\bf 4}, 339-349.

\bibitem{Harvey} Harvey, A.C., 1993,
long-memory in stochastic volatility.
Working paper (London School of Economics, London).

\bibitem{Hasbrouck91} Hasbrouck, J., Measuring the information content
  of stock trades, {\it J. Finance} {\bf 46}, 179-207 (1991).

\bibitem{hassler} Hassler, U. and J. Wolters, 1995,
Long-memory in inflation rates: International evidence.
{\it Journal of Business and Economic Statistics} {\bf 13}, 37-45.

\bibitem{Hausman92} Hausman, J.A. and A.W. Lo, 1992,
An ordered probitanalysis of transaction stock prices.
 {\it Journal of Financial Economics} {\bf 31} 319-379.

\bibitem{hosking} Hosking, J.R.M., 1981,
Fractional differencing.
{\it Biometrika} {\bf 68}, 165-176.

\bibitem{hurst} Hurst, H., 1951,
Long Term Storage Capacity of Reservoirs.
{\it Transactions of the American Society of Civil Engineers}
{\bf 116}, 770-799.

\bibitem{Kempf98} Kempf, A. and O. Korn, 1999,
 Market depth and order size. 
 {\it Journal of Financial Markets}, {\bf 2} 29-48 .
 
 \bibitem{leeready} Lee, C. and M. Ready, 1991,
Inferring Trade Direction from Intraday Data.
{\it Journal of Finance} {\bf 46}, 733-746.

\bibitem{lillo} Lillo, F.,  J.D. Farmer, and  R.N. Mantegna, 2003
Master curve for price impact function.
{\it Nature} {\bf 421}, 129-130.

\bibitem{agents} Lillo, F. {\it et al.} Agents in the market 
(in preparation).

\bibitem{lo}  Lo, A.W., 1991,
Long-term memory in stock market prices.
{\it Econometrica} {\bf 59}, 1279-1313.

\bibitem{lobato} Lobato, I.N. and C. Velasco, 2000,
Long-Memory in Stock-Market Trading Volume.
{\it Journal of Business \& Economic Statistics} {\bf 18}, 410-427.

\bibitem{mandelbrot0} Mandelbrot, B.B., 1971,
When Can Price Be Arbitraged Efficiently? A Limit to the 
Validity of the Random Walk and Martingale Models.
{\it Review of Economics and Statistics} {\bf 53}, 225-236.

\bibitem{mandelbrot1} Mandelbrot, B.B., 1972,
Statistical Methodology for Non-Periodic Cycles: From the Covariance
to R/S Analysis. 
{\it Annals of Economic and Social Measurement} {\bf 1}, 259-290. 

\bibitem{mandelbrot2} Mandelbrot, B.B., 1975,
Limit Theorems on the Self-Normalized Range for Weakly and Strongly
Dependent Processes.
{\it Z. Wahrscheinlichkeitstheorie verw. gebiete} {\bf 31}, 271-285. 

\bibitem{vanness} Mandelbrot, B.B. and J.W. van Ness, 1968,
Fractional Brownian motions, fractional noises and applications.
{\it SIAM Review} {\bf 10, No.4}, 422-437.

\bibitem{wealth} Pareto, V., 1896,
{\it Cours d'\'economie politique}.
Reprinted as a volume of {\it Oeuvres Complet\'es}
(Droz, Geneva, 1896-1965).

\bibitem{dfa} Peng, C.-K., S.V. Buldyrev, S. Havlin, M. Simons, and
  H.E. Stanley, 1994, Goldberger AL.,{ Mosaic organization of DNA
    nucleotides} {\it Physical Review E}{\bf 49},1685-1689.

\bibitem{Plerou01} Plerou, V., P. Gopikrishnan, X. Gabaix, and
  H.E. Stanley, 2002, Quantifying stock price response to demand
  fluctuations.  {\it Physical Review E} {\bf 66} 027104.

\bibitem{bouchaudimpact} Potters, M. and J.-P. Bouchaud, 2002,
More statistical properties of order books and price impact.
{\it Physica A} {\bf 324}, 133-140.

\bibitem{shea} Shea, G.S., 1991 Uncertainty and implied variance
  bounds in long-memory models of the interest rate term structure.
  {\it Empirical Economics} {\bf 16}, 287-312.

\bibitem{taqqu3} Taqqu, M.S., V. Teverovsky, and W. Willinger, 1995,
Estimators for long-range dependence: an empirical study.
{\it Fractals} {\bf 3} 785-788.

\bibitem{taqqu2} Teverovsky, V., M.S. Taqqu, and W. Willinger, 1999,
A critical look to Lo's modified R/S statistics.
{\it Journal of Statistical Planning and Inference} {\bf 80}, 211-227.

\bibitem{Torre97} Torre, N., 1997,
 {\it BARRA Market Impact Model Handbook},
BARRA Inc, Berkeley CA, www.barra.com.

\bibitem{taqqu1} Willinger, W., M.S. Taqqu, and V. Teverovsky, 1999,
Stock market prices and long-range dependence.
{\it Finance and Stochastics} {\bf 3}, 1-13.

\end{thebibliography}
\end{document}